\documentclass[12pt]{JHEP} 
\newcommand{\bexxx}{\begin{equation}}
\newcommand{\eexxx}{\end{equation}}
\newcommand{\beaxxx}{\begin{eqnarray}}
\newcommand{\eexxxa}{\end{eqnarray}}

\newcommand{\pint}{\makebox[0pt][l]{\hspace{2.4pt}$-$}\int}
\newcommand{\ellK}{{\rm K}}

\newcommand{\ellE}{{\rm E}}
\newcommand{\cJ}{{\cal J}}
\newcommand{\cE}{{\cal E}}
\newcommand{\tsmatthias}{\widetilde\sigma}
\newcommand{\sn}{\mathop{\mathrm{sn}}\nolimits}
\newcommand{\cn}{\mathop{\mathrm{cn}}\nolimits}
\newcommand{\dn}{\mathop{\mathrm{dn}}\nolimits}

\title{Matching Higher Conserved Charges for Strings and Spins}

\author{Gleb Arutyunov and Matthias Staudacher \\
Max-Planck-Institut
f\"ur Gravitationsphysik,
Albert-Einstein-Institut\\
Am M\"uhlenberg 1, D-14476 Golm, Germany\\
Email: \email{agleb,matthias@aei.mpg.de}}

\preprint{\hepth{0310182}\\
AEI-2003-080\\}

\abstract{We demonstrate that the recently found agreement between one-loop
scaling dimensions of large dimension operators in ${\cal N}=4$
gauge theory and energies of spinning strings on $AdS_5 \times S^5$ extends
to the eigenvalues of an {\it infinite} number of
hidden higher commuting charges.
This dynamical agreement is of a mathematically highly intricate
and non-trivial nature.
In particular, on the gauge side the generating function for
the commuting charges is obtained by integrable {\it quantum} 
spin chain techniques
from the thermodynamic density distribution function of Bethe roots.
On the string side the generating function, containing information
to arbitrary loop order, is constructed by solving exactly the
B\"acklund equations of the integrable {\it classical} string sigma model.
Our finding should be an important step towards
matching the integrable structures on the string and gauge side of the
AdS/CFT correspondence.
}


\begin{document}

\section{Introduction and summary}
Recently novel quantitative, dynamical tests of the duality between
superconformal ${\cal N}=4$ Yang-Mills
theory and type IIB superstrings on the $AdS_5\times S^5$
space-time background have been performed
\cite{ft2}-\cite{BFST}. They go beyond the BMN and the plane-wave
approximations to, respectively, gauge and string theories \cite{BMN}, and rely
on two crucial observations. The first one is the existence of a
sector of string states \cite{ft2,ft4,ft0,ft3} which is accessible
by semiclassical \cite{deVega:1994yz,GKP} methods, {\it i.e.}~one avoids the
unsolved problem of string quantization on a curved background.
The second observation is that planar ${\cal N}=4$ super
Yang-Mills theory is integrable in the one-loop approximation,
where it can be mapped to an integrable (super) spin chain
\cite{mz1,bes1,bes2}\footnote{ It has been conjectured that the
integrability extends to all loops \cite{bes3}.}. By using the
powerful method of the algebraic Bethe ansatz the latter property
allows one to find the exact one-loop anomalous dimensions of
arbitrary local composite operators in the gauge theory and
compare them with energies of the corresponding dual
semiclassical string states.

Specifically,
on the string theory side one looks for solitonic solutions
of the {\it classical} string sigma model with finite energy and spins
(i.e.~the commuting global charges of the SO(4,2)$\times$ SO(6) isometry group
of the $AdS_5\times S^5$ background).
With a natural spinning ansatz for the solution the sigma model
evolution equations are reduced to that of the Neumann integrable
system \cite{AFRT}. The latter have solutions of two different
types corresponding to so-called folded and circular strings.
Quite remarkably, in the thermodynamic limit the Bethe equations
encoding the one-loop anomalous dimensions of the gauge theory
operators exhibit the same types of solutions \cite{mz2}.
Moreover, one can show that in the limit of large spins\footnote{Possible 
connections with the tensionless limit of string theory and gauge theory 
multiplet shortenings have recently been discussed in \cite{MT}.} the first
subleading (order $\lambda$) term in the string energy, which is
a transcendental, parametric function of the spins, precisely
agrees with the one-loop anomalous dimension of the corresponding
operators both for the folded \cite{mz2,ft4} and for the circular
\cite{AFRT} configuration type.

It is clear that the Neumann system and the spin chain exhibit
very different types of integrability.
The first, striking difference is that one compares
the {\it classical} Neumann system to a {\it quantum} 
spin chain\footnote{
See \cite{Gorsky} for some ideas based on relation between the
discretized Neumann system and the XYZ spin chain.}.
The second difference is that the Neumann model has a {\it finite}
number (actually, just two) of integrals of motion while the 
spin chain exhibits, in the thermodynamical limit, an {\it infinite}
set of local, commuting integrals of motion. 
In both systems there are thus precisely as many integrals of motion 
as required in order to allow for an exact solution.
For the Neumann model a certain linear combination of the two integrals
gives the space-time energy of the string.
The model describes rigid strings (whose shape is time-independent)
and the ultimate role of the two integrals  appears in a rather
mild fashion:
they merely specify the topology of the solution (folded or circular).
In gauge theory these different string topologies are
reflected in the way the Bethe roots are distributed in the complex
plane. The energy is given by the lowest charge, but finding it
appears to require simultaneously diagonalizing all charges.
This is precisely what the Bethe ansatz achieves.
But what is the meaning of this infinite set of charges in the
framework of the Neumann model? 
It is clearly important to ask whether one can find them
in the dual string theory description as well. 
Answering this question is the main goal of this paper.

Before we proceed let us mention that the possible interrelation
between the integrability
of the string theory on the $AdS_5\times S^5$ background and
its field theory counterpart has been recently discussed from the viewpoint
of nonabelian (nonlocal) symmetries generated by Yangians 
\cite{MSW,BPR,DNW} (see also \cite{Al} for further developments).
We feel however that it is of primary importance to first understand the map
between the infinite families of local commuting integrals of motion,
whose existence is crucial for the solubility of the gauge and string
theories at hand.

Actually it is well-known, starting from the work of
Pohlmeyer \cite{Pol}, that the classical principle
sigma models and their various reductions
(in particular the O(n) vector models we are interested in here)
possess an infinite family of local commuting charges
\cite{Cherednik:qj, Ogielski:hv,DeVega:1992xc}. The easiest
way, perhaps, to exhibit this family is to use the so-called
B\"acklund transformation.

Let us now come to the description of the present work and the
results obtained.
Our consideration starts from the one-loop gauge theory.
We first construct the generating function (resolvent) of the
infinite family of commuting local spin chain charges $Q_n$,
as obtained from the Bethe solutions corresponding to
folded and circular string configurations.
The first non-trivial charge $Q_2$, obtained already previously in
\cite{mz2,BFST},
is the eigenvalue of the one-loop dilatation operator, {\it i.e.}~it
gives the anomalous dimension. An intriguing feature of the family of
gauge theory charges is that, if properly defined, they satisfy ``BMN
scaling'' \cite{BMN}: The $n$-th charge $Q_n$ scales as $J^{1-n}$,
where $J$ is the total spin (i.e.~the length of the chain).

Next, using the B\"acklund transformation
for the O(6) model, we explicitly work out the first few
commuting charges ${\cal E}_n$ in string theory and further evaluate
them on the folded and circular
type solutions of the Neumann system. Here the first charge
appears to coincide with the Hamiltonian of the Neumann
system which, due to the Virasoro constraint, gives
the energy of the string.
In both the gauge and the string theory computations
the charges emerge as rather nontrivial combinations
of elliptic integrals governed by the gauge/string modular parameter.
In each case the modular parameter is determined by a transcendental
relation expressing it as a function of the spins. The string theory
calculation comprises the all-loop result for the higher charges.
In order to compare with the one-loop gauge theory predictions
we develop a perturbative expansion of ${\cal E}_n$.
The obtained ``one-loop'' charges predicted by string theory
do not yet satisfy the above mentioned property of BMN scaling.
This is shown to be due to mixing effects,
and BMN scaling can be recovered after some natural linear redefinitions
${\cal E}_n\to {\cal Q}_n$.
Finally, we demonstrate that a Gauss-Landen modular transformation
maps the entire family of commuting charges
of string theory onto the one of the gauge theory!

This result nicely extends the correspondence picture: not only the
energy of a string state matches the anomalous dimension but
also the higher string commuting charges can be matched
to those of the gauge theory.
Although we have demonstrated this matching on some particular string/gauge
solutions it is clear that all this hints at a universal relation
between the string sigma model and the spin chain: Semiclassically
the string sigma model
and the spin chain are the same.
The Neumann solutions and their gauge theory counterparts
are just distinguished from the ``sea'' of all possible solutions
by specifying the values of the infinite number of local commuting charges.
Further evidence for this universality comes from our finding that
the just mentioned linear redefinitions do not depend on the
solution type. 

The knowledge of the whole
family of the string commuting charges relies essentially on
on the knowledge of the solution of the  B\"acklund equations.
As we will see the perturbative
solution can be pushed to any desired order in the spectral parameter $\gamma$ but it does not
allow us to obtain all the charges; an exact, i.e. non-perturbative 
in $\gamma$,
solution of the  B\"acklund equations is needed.
Fortunately, the problem of finding an exact solution
turns out to have a beautiful resolution.
The perturbative
treatment shows that the B\"acklund solution still
solves the evolution equations of the Neumann model.
On the other hand we
know the most general solution of the Neumann model;
for the two-spin case it depends on one arbitrary
constant $\nu$, which can be assumed to depend
on the spectral parameter $\gamma$. The B\"acklund equations
are then used to determine the function $\nu\equiv\nu(\gamma)$.
In this way we get the explicit spectral-dependent
solution and use it to obtain the exact generating function
(string spectral curve) $\cE(\gamma)$ of
commuting charges in string theory. The final expression for
$\cE(\gamma)$ turns out to be rather simple. However, it
generates unimproved charges, i.e.~the charges 
do not yet obey BMN scaling.
As was already mentioned, the charges with the ``proper''
one-loop BMN scaling can be obtained from the
original unimproved charges by a certain linear redefinition,
as we verified explicitly to high order.
This does not come as a surprise: Within the commuting
family there is always freedom to apply arbitrary linear
transformations on the complete set of charges.
Using this freedom we are able to derive the {\it complete}
generating function of higher one-loop gauge theory charges 
from string theory! 

The upshot of the present work is then that we shed light 
on matching the integrable structures on the gauge and string
sides of the AdS/CFT correspondence. This is done by
way of example, for two specific cases. It would
be very important to find the underlying general reasons
for our findings. The structures on the gauge and string
sides are still conceptually rather distinct. E.g.~on the gauge
side the commuting charges are actually a crucial device,
as they appear to be needed to find the energies; on the other hand, on the
string side we can find the energies without requiring
the knowledge of the higher charges: here they emerge,
at this point, as a mere add-on. The reason for that  can be traced to 
our assumption of the Neumann (rotating) ansatz 
which considerably restricts the possible string trajectories. 
To end, 
we quite generally feel that the local commuting charges 
might well hold the key to unraveling the mysterious integrable structures 
of the planar AdS/CFT system.

\medskip
The paper is organized as follows.
In chapter 2 we explain how to derive the generating function
of commuting charges in one-loop gauge theory,
first quite generally, and subsequently in the specific case of
the ``folded'' and ``circular'' root distributions.
This is done by inspecting the detailed statistical distribution
of Bethe roots in the integrable quantum spin chain approach.
In chapter 3 we first discuss
the classical string solutions within the Neumann integrable ansatz.
Then we describe
a perturbative construction of the B\"acklund transformations and
subsequently use it
to compute the first few commuting charges in string theory. We
furthermore show that in the one-loop approximation these
charges can be precisely matched to those
in the gauge theory. Then we show how the perturbative construction
of the B\"acklund transformation allows one
to guess an ansatz for an exact solution of the B\"acklund
equations. In appendix A the  B\"acklund equations are solved
exactly and the solution is used in chapter 3 to construct the
complete generating function for commuting charges in string theory.
As mentioned above this function generates unimproved charges.
We therefore demonstrate in the final chapter 4 how to extract
the complete infinite set of one-loop gauge charges from the
string generating function. 
To keep the presentation more transparent,
we restrict ourselves in the main text devoted to the string
charges to the case of the folded
string; the circular string is treated separately in appendix A.
In appendix A we also discuss in some detail the
modular Gauss-Landen transformation which relates
the gauge and the string theory quantities and explain its
geometric meaning in our present context.
In appendix B our conventions for elliptic integrals and elliptic
functions are summarized.

\section{Higher one-loop charges in gauge theory}

\subsection{Generalities}

Let us continue the study of the spectrum of the following
single trace ${\cal N}=4$ gauge theory operators which were shown
to be dual to $AdS$ strings rotating on the five sphere $S^5$ in
two orthogonal planes with large angular momenta
$J_1$,$J_2$ \cite{mz2,ft4,BFST}:
\bexxx
\label{eq:embryo}
\mathop{\mathrm{Tr}} Z^{J_1} \Phi^{J_2}  + \ldots \ ,
\eexxx
These operators are composites of two ($Z,\Phi$) out of the three complex
scalar fields of the ${\cal N}=4$ gauge theory.
The dots in eq.(\ref{eq:embryo})
indicate that the operator mixes in the $N=\infty$ planar gauge theory
with all single trace operators containing arbitrary sequences
of $J_1$ $Z$ and $J_2$ $\Phi$ fields. We therefore need to diagonalize,
i.e.~find the eigensystem, in the space of all these operators. This is a
formidable problem even at the one-loop level; it  is solved by
mapping it to the spectral problem of an equivalent
XXX Heisenberg quantum spin chain \cite{mz1}, where the two fields are
interpreted as, respectively, up-spins and down-spins.
The exact planar one-loop spectrum $\frac{1}{4 \pi^2} Q_2$
of these operators is then parametrically determined \cite{mz1}
by the expression
\bexxx
\label{eq:discreteQ2}
Q_2=\frac{J}{2}\sum_{j=1}^{J_2}
\frac{1}{u_j^2+{1/4}}
\eexxx
where the parameters $u_j$ (the Bethe roots) are found by solving
the following set of Bethe equations for all $J_1 \geq J_2$,
where $J=J_1+J_2$:
\bexxx
\label{eq:Bethe}
\left( \frac{u_j+i/2}{u_j-i/2} \right)^J=
\prod_{{k=1} \atop {k\neq j}}^{J_2}
\frac{u_j-u_k+i}{u_j-u_k-i}\, \qquad \qquad
\prod_{j=1}^{J_2} \frac{u_j+i/2}{u_j-i/2}=1.
\eexxx
No two roots are allowed to coincide.
A similar but obviously more involved situation applies to
arbitrary composite operators in ${\cal N}=4$ gauge theory,
where the Bethe equations are also known \cite{bes2}.

We are interested in the thermodynamic limit
$J,J_2 \rightarrow \infty$ of these Bethe equations with
\bexxx\label{eq:filling}
\alpha \equiv \frac{J_2}{J}
\eexxx
held fixed.
The Bethe roots are expected to condense onto a union
${\cal C}=\cup_n {\cal C}_n$  of smooth contours
${\cal C}_n$ in the complex plane of the continuous variable $u$:
\bexxx
\label{eq:u}
\frac{u_j}{J} \rightarrow u.
\eexxx
Their distribution may be described by a density $\rho(u)$ having
support $u \in {\cal C}$ on the union of contours:
\bexxx
\label{eq:smooth}
\rho(u)=\frac{1}{J} \sum_{j=1}^{J_2} \delta
\left(u-\frac{u_j}{J}\right),
\quad {\rm hence} \quad
\int_{\cal{C}} du~\rho(u) = \alpha.
\eexxx
The density allows to find the energy eq.(\ref{eq:discreteQ2})
in the thermodynamic limit.
\bexxx
\label{eq:continuumQ2}
Q_2=\frac{1}{2} \int_{\cal C} du~\frac{\rho(u)}{u^2}.
\eexxx
The discrete Bethe equations eqs.(\ref{eq:Bethe}) turn into
\bexxx
\label{eq:BetheCont}
\pint_{\mathcal{C}} dv~
\frac{ \rho(v)\, u}{v-u}=\frac{1}{2}+\pi \,
n_{\mathcal{C}(u)} \,u,\quad
{\rm where} \quad u \in {\cal C}_n \quad {\rm and} \quad
\int_{\mathcal{C}}\,du~\frac{ \rho(u)}{u}=0.
\eexxx
This system of singular integral equations (the bar through the integral
indicates a principal part prescription) determines the density.
The mode numbers $n_{\mathcal{C}(u)}$ are integers which are
expected to be constant on each smooth component ${\cal C}_n$ of the
density support ${\cal C}$. They appear since we need to take
a logarithm of the discrete equations (\ref{eq:Bethe}) in order to
derive eqs.(\ref{eq:BetheCont}), and are subject to the condition
$\int_{\mathcal{C}} du\, \rho(u)\, n_{\mathcal{C}(u)} = 0$.

The solutions of the system of equations eqs.(\ref{eq:BetheCont}) have not yet
been studied in any generality. Such a study would presumably
lead to a classification of the large $J_1,J_2$ spectrum of the operators
(\ref{eq:embryo}). In general we expect solutions for any number of
cuts ${\cal C}_i$. An even larger system of singular integral
equations, which has not even yet been written down in public,
governs the full set
of Bethe equations \cite{bes2} for arbitrary single trace operators
in the thermodynamic limit.
What has been done was a study of the simplest
kinds of solution of eqs.(\ref{eq:BetheCont})
involving one or two contours \cite{mz2,BFST}.
These solutions are
reviewed and further analyzed in sections 2.2. and 2.3.

At this point we should realize that the above solution
of the ``thermodynamic'' spectral problem of the operators
eq.(\ref{eq:embryo}) contains
a lot more information than we have used so far. The
Bethe integral equations eqs.(\ref{eq:BetheCont}) allow to determine
the root distribution density $\rho(u)$ which is
then merely used to extract the energy eigenvalue
eq.(\ref{eq:continuumQ2}).

E.g.~the so far obtained solutions \cite{mz2,BFST} found for
$\rho(u)$ an elliptic curve, ubiquitously present in integrable
systems. The resulting charge $Q_2$ was successfully matched to
a string computation \cite{ft4,BFST}.
It is thus very natural to ask whether the entire ``gauge curve''
$\rho(u)$ also has an interpretation on the string side.
We will find that the answer to this question is: Yes indeed!
In fact, solving the Bethe ansatz equations eqs.(\ref{eq:Bethe})
leads not only to the diagonalization of the Hamiltonian $Q_2$ (=lowest
charge) of the spin chain, but also simultaneously diagonalizes the
entire set of commuting charges $Q_k$. These are generated by
the transfer matrix $T(u)$ of the discrete spin chain through
\bexxx
\label{eq:chargedef}
Q_{k+1} = -\frac{i}{2} \frac{ J^k}{k!}~
\frac{d^k}{d u^{k}} \log T(u) \Bigg|_{u=\frac{i}{2}}
\eexxx
Each Bethe state, determined through eqs.(\ref{eq:Bethe}), is
an eigenstate of $T(u)$ for arbitrary complex values of the
spectral parameter $u$. Its eigenvalue is
\bexxx
\label{eq:transfer}
T(u)=
\left(u+\frac{i}{2}\right)^J \prod_{j=1}^{J_2}
\frac{u-u_j-i}{u-u_j}+
\left(u-\frac{i}{2}\right)^J \prod_{j=1}^{J_2}
\frac{u-u_j+i}{u-u_j}\, .
\eexxx
Dropping an irrelevant $u_j$-independent constant, we find in the
thermo\-dynamic limit (\ref{eq:u}),(\ref{eq:smooth}) for the charges
\bexxx\label{eq:allcharges}
Q_{k}=\frac{1}{2}\int_{\cal C}du~\frac{\rho(u)}{u^{k}} \ .
\eexxx
Clearly the commuting charges are non-unique up to linear (or even polynomial)
redefinitions. Our definition eq.(\ref{eq:chargedef}),
which is standard except for the factor $J^k$, ensures
that the charges behave properly (i.e.~they are of order
${\cal O}(1)$) in the scaling limit (\ref{eq:u}),(\ref{eq:smooth}).
It is natural to introduce a generating function of all conserved
charges  $Q_{k}$ (the resolvent). It is given by
\bexxx\label{eq:resolventdef}
H(u)=\frac{\alpha}{2}+\frac{1}{2} \sum_{k=1}^{\infty}~Q_{k}~u^{k},
\qquad {\rm i.e.} \qquad
H(u)=\frac{1}{2} \int_{{\cal C}} dv~\rho(v) \frac{v}{v-u}
\eexxx
and thus analytically defined throughout the complex $u$ plane,
except for the cuts ${\cal C}_i$. Let us now, for the reminder of
this chapter, compute the resolvent for the known cases of the
``folded'' and ``circular'' cases and thereby derive the infinite
number of commuting charges for these solutions. In the next
chapter these one-loop charges will be reproduced on the string
side.

\subsection{Folded case}

Let us first reconsider the large $J_1,J_2$ ground state energy of
the operators (\ref{eq:embryo}) \cite{mz2,BFST}. In this case the
roots condense on two curved, convex cuts ${\cal C}={\cal C}^+
\cup {\cal C}^-$ which are symmetric w.r.t~reflection on both the
real and imaginary axis. In order to have explicit formulas, we
will again discuss the related case where the two cuts (${\cal
C}^- \rightarrow [-b,-a]$ and ${\cal C}^+ \rightarrow[a,b]$) are
located on the real axis in the $u$-plane: This corresponds to
analytically continuing the filling fraction $\alpha=J_2/J$ to
negative values \cite{mz2,BFST}. The thermodynamic Bethe equations become  
\bexxx\label{eq:airfoil}
\pint_a^b dv~\frac{\rho(v)\,u^2}{v^2-u^2}
=\frac{1}{4}-\frac{\pi}{2}\,u\ ,
\qquad
Q_2=\int_a^b du~\frac{\rho(u)}{u^2} \ .
\eexxx
The first equation is closely related to the saddle point equation of the 
O($\pm 2$) matrix model (see e.g.~\cite{KS}).
Physically this corresponds to a string rotating with one spin
on $AdS_5$ and one angular momentum on $S^5$ \cite{BFST},
but we will not make use of this fact in this paper.
The previously obtained solution reads
\bexxx\label{eq:denssol}
\rho(u)=\frac{2 }{\pi u} \pint_a^b dv~
\frac{v^2}{v^2-u^2}
\sqrt{\frac{(b^2-u^2)(u^2-a^2)}{(b^2-v^2)(v^2-a^2)}}\,.
\eexxx
with the normalization (remember $\alpha<0$, i.e.~formally $J<0$)
\bexxx\label{eq:norm}
\int_a^b du~\rho(u)=-\frac{\alpha}{2}\, .
\eexxx
The interval boundaries are determined by
\bexxx\label{eq:ab}
a=\frac{1}{4 \ellK(q)},
\qquad
b=\frac{1}{4 \sqrt{1-q}~\ellK(q)}\, .
\eexxx
Here $\ellK(q)$ is the complete elliptic integral of the 
first kind (see appendix B on our conventions for elliptic integrals and 
elliptic functions).
The density may be expressed explicitly through the elliptic integral
of the third kind:
\bexxx\label{eq:bigpi}
\rho(u)=\frac{1}{2 \pi b u}
\sqrt{\frac{u^2-a^2}{b^2-u^2}}
\left[\frac{b^2}{a}-4 u^2 \Pi\left(\frac{b^2-u^2}{b^2},q\right) \right],
\qquad
q=\frac{b^2-a^2}{b^2},
\eexxx
{}From this it is straightforward to obtain the energy eigenvalue
$Q_2$ in the parametric form
\bexxx\label{eq:Q2}
Q_2=2 \,
\ellK(q)\Big[ (2-q)\ellK(q)-2\ellE(q) \Big]\, ,
\eexxx
with
\bexxx\label{eq:modulus}
-\alpha\equiv  -\frac{J_2}{J}
=\frac{1}{2\sqrt{1-q}}\,\frac{\ellE(q)}{\ellK(q)}-\frac{1}{2}  \ .
\eexxx
This rather involved final result, first obtained in
\cite{mz2,BFST} agrees with the one-loop prediction of
semiclassical string theory \cite{ft4,BFST}. This ends our short
review and we will now compute the higher charges, using the
procedure explained in the previous section.

For the even charges
(the odd charges vanish on the unpaired ground state we are interested
in) we have
\bexxx\label{eq:allcharges1}
Q_{2 k}=\int_a^b du~\frac{\rho(u)}{u^{2 k}} \ .
\eexxx
\bexxx\label{eq:resolventdef1}
H(u)=-\frac{\alpha}{2}+\sum_{k=1}^{\infty}~Q_{2 k}~u^{2 k},
\qquad {\rm i.e.} \qquad
H(u)=\int_a^b dv \rho(v)~\frac{v^2}{v^2-u^2} .
\eexxx
On the cut $u \in \left[a,b\right]$
this function behaves as
$H(u \pm i \epsilon)=-\frac{\alpha}{2}+\frac{1}{4} -\frac{\pi}{2} u \pm
i \frac{\pi}{2} u \rho(u)$.
It is therefore straightforward to obtain an integral representation of the
resolvent
\bexxx\label{eq:resolvent1}
H(u)=-\frac{\alpha}{2}+\frac{1}{4}-
\int_a^b \frac{dv\, v^2}{v^2-u^2}
\sqrt{\frac{(b^2-u^2)(a^2-u^2)}{(b^2-v^2)(v^2-a^2)}}.
\eexxx
Just like the density eq.(\ref{eq:bigpi}) we can also express it explicitly with
the help of the elliptic integral of the third kind:
\bexxx\label{eq:altresolvent}
H(u)=-\frac{\alpha}{2}+\frac{1}{4} -\frac{\pi}{2} u-
\frac{1}{4 b}
\sqrt{\frac{a^2-u^2}{b^2-u^2}}
\left[\frac{b^2}{a}-4 u^2 \Pi\left(\frac{b^2-u^2}{b^2},q\right) \right].
\eexxx
This is not the most convenient form to perform the power series
expansion around $u=0$. A further, more suitable representation for
$H(u)$ is
%
%
%
\bexxx\label{eq:resolvent2r}
H(u)=-\frac{\alpha}{2}+\frac{1}{4}
-\frac{a^2}{b} \sqrt{\frac{b^2-u^2}{a^2-u^2}}
~\Pi\left(-q\frac{u^2}{a^2-u^2},q\right).
\eexxx
{}From this expression we can easily obtain the
explicit form of the higher charges; the first few read
{\small
\beaxxx\label{eq:Q4}
Q_4~&=&\frac{2^3}{3}
\ellK(q)^3 \Big[ 4 (q-2) \ellE(q)+(8-8 q+3 q^2) \ellK(q) \Big]\, , \\
\nonumber
Q_6~&=&\frac{2^6}{5}\ellK(q)^5
\Big[ -2 (8-8q+3q^2) \ellE(q)+(16-24q+18 q^2-5q^3) \ellK(q) \Big]\, ,\\
\nonumber
Q_8~&=&\frac{2^7}{7}\ellK(q)^7
\Big[8(q-2)(8-8q+5q^2) \ellE(q)+(128-256q+288q^2-160q^3+35q^4) \ellK(q) \Big]\, ,\\
\nonumber
Q_{10}&=&\frac{2^{10}}{9}\ellK(q)^9
\Big[(-256+512q-576q^2+320q^3-70q^4)
\ellE(q)  \\ \nonumber
&&~~~~~~~~~~~~~~~-(q-2)(128-256q+352q^2-224q^3+63q^4) \ellK(q) \Big]\,
,\\
\vdots& & 
\nonumber
\eexxxa
}
In the next chapter we will demonstrate how to ``perturbatively'' (in $u$)
find these charges on the string side.
In the final chapter 4 we will then rigorously derive the entire 
generating function eq.(\ref{eq:resolvent2r}) from string theory.
However, before doing so we will also
treat the Bethe solution corresponding to the circular string.

\subsection{Circular case}

The configuration of Bethe roots corresponding to the circular string
was found in \cite{mz2}, chapter 4 (see also \cite{BFST}, appendix D).
As opposed to the folded case, this is not the ground state of
the spin chain.
For the convenience of the reader we will closely follow\footnote{
However, in order (1) to avoid confusions with the notation in other
parts of the present paper, and (2) to agree with the perhaps more
natural normalization of the folded string,
we have decided to make the following
notational changes w.r.t.~\cite{mz2,BFST}:
(1) $q \rightarrow x$, $q' \rightarrow y$,
$s \rightarrow c$, $t \rightarrow d$
and (2) $\sigma(q) \rightarrow 2 \sigma(x)$.}
\cite{mz2,BFST}. In the present case the roots are
distributed on a single, connected curve ${\cal C}$
along the imaginary axis. Around the origin
an odd number $J_2'$ (with $J_2'<J_2\leq J_1$ and $J=J_1+J_2$) of roots
are placed in the near vicinity of the half-integer purely imaginary values,
i.e.~their density is constant ($=2$ since there are two roots
per integer interval on the imaginary axis), up to $\pm i c$ with
$c= \frac{J_2'-1}{4 J}$.
Further away the distribution is less dense (i.e.~$\tsmatthias(x) < 2$)
and non-constant.
Let us replace the variable $u$ parametrizing the Bethe plane by $u=i x$.
The density $\sigma(x)\equiv \rho(u)$ then has the form
\bexxx\label{eq:rootden}
\sigma(x)=\frac{1}{J} \sum_{j=1}^{J_2} \delta \left(x+i
  \frac{u_j}{J}\right)
\quad \rightarrow \quad
\sigma(x)\ =\ \left\{ \begin{array}{ll}  2 &\qquad -c<x<c, \\
\tsmatthias(x) &\qquad  c<x<d, \\ \tsmatthias(-x)  &
\qquad -d<x<-c, \\     0
&\qquad x<-d\ \ {\rm or}\ \ d<x \end{array} \right.
\eexxx
Due to the constant condensate around the origin we again end up
with a two-cut problem, despite the fact that all roots
lie on one continuous curve ${\cal C}$!
The continuum Bethe equations and, respectively, the lowest
charge (i.e.~energy) read
\bexxx\label{eq:airfoil2}
\pint_c^d dy\,\tsmatthias(y)\, \frac{x^2}{x^2-y^2}
=\frac{1}{4}- x \log \frac{x+c}{x-c}
\qquad
Q_2=\frac{2}{c} - \int_c^d dx~\frac{\tsmatthias(x)}{x^2} \ .
\eexxx
The solution is
\bexxx\label{eq:denssol2}
\tsmatthias(x)=\frac{1}{2 \pi} \sqrt{(d^2-x^2)(x^2-c^2)}
\left( -\frac{1}{x c d}+
4 \int_{-c}^c
\frac{dy}{x-y}
\frac{1}{\sqrt{(d^2-y^2)(c^2-y^2)}} \right),
\eexxx
with the normalization
\bexxx\label{eq:norm2}
2 c+\int_c^d dx~\tsmatthias(x)=\frac{\alpha}{2}.
\eexxx
The interval boundaries are determined through
\bexxx\label{eq:cd}
c=\frac{1}{8 \ellK(r)},
\qquad
d=\frac{1}{8 \sqrt{r}~\ellK(r)}.
\eexxx
The density may again be expressed explicitly
\bexxx
\label{eq:bigpi2}
\tsmatthias(x)=2-\frac{4}{\pi d x} \sqrt{(d^2-x^2)(x^2-c^2)}
~\Pi \left(\frac{x^2}{d^2},r\right),
\qquad
r=\frac{c^2}{d^2}.
\eexxx
%
%
%
The energy eigenvalue is, similar to, but different from
eq.(\ref{eq:Q2}), parametrically given by
\bexxx
\label{eq:Q2-2}
Q_2=
8 \ellK(r)\Big[2\ellE(r)+(r-1)\ellK(r)\Big]
\eexxx
with
\bexxx
\label{eq:modulus2}
\alpha\equiv  \frac{J_2}{J}
=\frac{\sqrt{r}-1}{2 \sqrt{r}}+\frac{1}{2 \sqrt{r}} \frac{\ellE(r)}{\ellK(r)}  \ .
\eexxx
These are the gauge results derived for the circular string in
\cite{mz2,BFST} and successfully compared to semiclassical string
theory in \cite{AFRT}.

As for the previous case of the folded solution we would now like
to also extract the higher charges from the density function
eq.(\ref{eq:bigpi2}). Naively these should be given,
{\it cf} eqs.(\ref{eq:allcharges}),(\ref{eq:allcharges1}), by the
expression
\bexxx\label{eq:allchargeswrong}
Q_{2 k}\stackrel{?}{=}
(-1)^k \int_0^d dx~\frac{\sigma(x)}{x^{2k}}.
\eexxx
However, this formula is clearly nonsensical: The density $\sigma(x)$
is constant at the origin, and the above expression therefore
diverges at $x=0$. Let us go back to the original definition of the
higher charges for the finite spin chain with discrete roots $u_j=i x_j$,
i.e.~eqs.(\ref{eq:chargedef}),(\ref{eq:transfer}), and
investigate the thermodynamic limit more carefully.
As we are assuming $J$ to be large (i.e.~the second term in the
transfer matrix eq.(\ref{eq:transfer}) does not contribute) we have
\bexxx
\label{eq:chargedef2}
Q_{2 k} = \frac{i^{2 k}}{2} \frac{ J^{2 k-1}}{(2k-1)!}~
\left[ \frac{d^{2k-1}}{d x^{2k-1}} \log
\prod_{j=1}^{J_2} \frac{x-x_j-1}{x-x_j}
\right]_{x=\frac{1}{2}} + \ldots
\eexxx
where the dots indicate an $x_j$-independent constant, which we can drop.
We had assumed that all roots $u_j=i x_j$ scale as $\sim J$.
This is clearly not the case for the roots near the origin.
However, these are condensed at half-integer values of the imaginary
axis, and their contribution to the transfer matrix mostly cancels,
see eq.(\ref{eq:chargedef2}). We therefore have
\bexxx
\label{eq:chargedef3}
Q_{2 k} = \frac{i^{2 k}}{2} \frac{ J^{2 k-1}}{(2k-1)!}~
\left[ \frac{d^{2k-1}}{d x^{2k-1}} \left(
\log \frac{(x-\frac{J_2'+3}{4})(x-\frac{J_2'+1}{4})}
{(x+\frac{J_2'-3}{4})(x+\frac{J_2'-1}{4})} +
\log \prod_j \frac{x-\tilde x_j-1}{x-\tilde x_j} \right)
\right]_{x=\frac{1}{2}}
\eexxx
where the last product runs only over the $J_2-J_2'$
roots $\tilde x_j$ located
outside the condensate, and distributed according to the density
$\tsmatthias(x)$. Taking the thermodynamic limit $J,J_2,J_2' \rightarrow \infty$
we find that the charges are actually perfectly finite and given by
\bexxx
\label{eq:allcharges2}
Q_{2 k}=\frac{2}{2 k-1} \frac{1}{c^{2 k-1}} - \int_c^d dx
\frac{\tsmatthias(x)}{x^{2 k}}.
\eexxx
(For esthetic reasons we have dropped an overall immaterial
sign $(-1)^{k+1}$ in this expression.)
Eq.(\ref{eq:allcharges2}) generalizes a result proven by different
means in \cite{mz2}
for the energy ($Q_2$) to arbitrary $2 k$. Interestingly,
eq.(\ref{eq:allcharges2}) coincides with an ad-hoc regularization
of the naive formula eq.(\ref{eq:allchargeswrong}), where
one just drops the divergence at the lower limit of the
integration. The generating function of the charges
eq.(\ref{eq:allcharges2}) (i.e.~the resolvent) is found to be
\bexxx
\label{eq:resolventdef2}
H(x)=\frac{\alpha}{2}-\sum_{k=1}^{\infty}~Q_{2 k}~x^{2k}
\quad {\rm i.e.} \quad
H(x)=2 c+ x \log \frac{c-x}{c+x} + \int_c^d dy \tsmatthias(y)
\frac{y^2}{y^2-x^2}
\eexxx
We can again express it explicitly with
the help of the elliptic integral of the third kind:
\bexxx\label{eq:resolvent2}
H(x)=\frac{\alpha}{2}-\frac{1}{4}
+\frac{2}{d} \sqrt{(d^2-x^2)(c^2-x^2)}~
\Pi \left(\frac{x^2}{d^2},r\right).
\eexxx
{}From these expressions we can easily obtain the explicit form of the
higher charges; the first few are
{\small
\beaxxx
\label{eq:Q4-2}
Q_4~&=&\frac{2^7}{3}\ellK(r)^3\Big[4(1+r)\ellE(r)
+(r-1)(3r+1)\ellK(r) \Big]\, , \\
\nonumber
Q_6~&=&\frac{2^{12}}{5}\ellK(r)^5\Big[(6+4r+6r^2)
\ellE(r)+(r-1)(1+2r+5r^2)\ellK(r)\Big]\, ,\\
\nonumber
Q_8~&=&\frac{2^{15}}{7}\ellK(r)^7\Big[8(r+1)(5-2r+5r^2)
\ellE(r)+(r-1)(5+9r+15r^2+35r^3)\ellK(r)\Big]\, ,\\
\nonumber
Q_{10}&=&\frac{2^{20}}{9}\ellK(r)^9\Big[(70+40r+36r^2+40r^3+70r^4)
\ellE(r) \\
\nonumber
&&~~~~~~~~~~~~~~~~~
+(r-1)(7+12r+18r^2+28r^3+63r^4)\ellK(r)\Big]\,,\\
\vdots& & \nonumber
\eexxxa
}
Using the methods that will be exposed in the next two chapters,
we will derive in appendix A.2 the charges (\ref{eq:Q4-2}), and actually
the entire generating function eq.(\ref{eq:resolvent2}) from string theory.

\section{Higher semiclassical charges in string theory}

\subsection{Classical string solutions}
The classical closed bosonic string propagating in the
$AdS_5\times S^5$ space-time is described in terms of the
\mbox{O(6)$\times$O(4,2)} sigma model. As was shown in
\cite{AFRT}, rigid string solitons (i.e.~strings with a
time-independent profile) are classified in terms of periodic
solutions of the Neumann integrable system. The Neumann system
arises from the sigma model equations of motion after specifying
a {\it rotating} string ansatz for the embedding
coordinates.\footnote {More precisely, one obtains two Neumann
systems: The ``compact'' system originating from the O(6) part and
the ``non-compact'' one related to the O(4,2) part. The conformal
gauge constraint imposes one equation between the conserved
energies of the systems.}  In particular, for the motion on $S^5$
the rotating ansatz reads \beaxxx
X_1+iX_2=x_1(\sigma)~e^{iw_1\tau}\, , ~~~
X_3+iX_4=x_2(\sigma)~e^{iw_2\tau}\, , ~~~
X_5+iX_6=x_3(\sigma)~e^{iw_3\tau}\, , \nonumber \eexxxa where $X_i$
are the embedding coordinates of $S^5$. The variables
$x_i(\sigma)$ satisfy the dynamical equations of the Neumann
system and the relation \beaxxx \sum_{i=1}^3x_i^2=1 \, , \eexxxa which
implies that at a fixed moment in time the string lies on a
two-dimensional sphere $S^2$. In addition the $x_i(\sigma)$ obey
the closed string periodicity condition: \beaxxx
x_i(\sigma+2\pi)=x_i(\sigma) \, . \eexxxa

The rotation on $S^5$ with the frequencies
$w_i$ gives rise to three angular momentum integrals (spins)
$J_i=\sqrt{\lambda}\cJ_i$. 
Due to the Virasoro constraint the
space-time energy of the string $E=\sqrt{\lambda}{\cal E}$
becomes a function of the spins:
\beaxxx
E=E(J_1,J_2,J_3)\, .
\eexxxa
{}From now on we will restrict our consideration to
string solitons carrying only two non-vanishing spin components,
$J_1$ and $J_2$. Needless to say, it would be quite interesting
to extend the analysis below to the general three-spin case,
using as an input \cite{AFRT}.

Recall that the
solutions for $x_i$ depend on a discrete choice for the
relative position of the two Neumann integrals
w.r.t.~to the frequencies $w_i^2$: these can be of two types, corresponding to
folded or circular string geometries. Below we will review
these solutions \cite{ft4,AFRT}.

\medskip

\noindent {\it Folded string solution}
\medskip

\noindent The embedding coordinates for the folded string configuration
are solved in terms of the standard Jacobi elliptic functions
(cf.~appendix B) as follows ($w_{ij}^2\equiv w_i^2-w_j^2$):
\beaxxx
\label{fs}
x_1(\sigma)=\mbox{dn}\Big(\sigma\sqrt{w_{21}^2},~ t\Big)\, , ~~~
x_2(\sigma)=\sqrt{t}~\mbox{sn}\Big(\sigma\sqrt{w_{21}^2} ,~ t\Big)\, , ~~~x_3(\sigma)=0\, ,
\eexxxa
Here the Jacobi modulus $t$ is determined through the 
closed string periodicity condition (for simplicity we consider 
the solution with the winding number $n=1$, cf.~the general discussion
of winding numbers in \cite{AFRT}):
\beaxxx
\label{cspcf}
\frac{\pi}{2}\sqrt{w_{21}^2}=\ellK(t) \, , ~~~~~w_2>w_1
\eexxxa 
and, solving for $w$'s in terms of spins $\cJ_i$, it can be further found from
the equation
\beaxxx
\label{Jmf}
\left(\frac{\cJ_2}{\ellK(t)-\ellE(t)}\right)^2-\left(\frac{\cJ_1}{\ellE(t)}\right)^2=\frac{4}{\pi^2} \, .
\eexxxa

\newpage

\medskip

\noindent {\it Circular string solution}
\medskip

\noindent The circular type configuration is described by
\beaxxx
\label{cs}
x_1(\sigma)=\mbox{sn}\Big(\sigma\sqrt{\frac{w_{12}^2}{t}},~ t\Big)\, ,
~~~x_2(\sigma)=\mbox{cn}\Big(\sigma\sqrt{\frac{w_{12}^2}{t}},~ t\Big)\, , ~~~x_3(\sigma)=0\, ,
\eexxxa
while the equation for the modulus (again for the winding number $n=1$) is
\beaxxx
\frac{\pi}{2}\sqrt{\frac{w_{12}^2}{t}}=\ellK(t) \, , ~~~~~w_1>w_2
\eexxxa
and it can be reexpressed via the spins as follows
\beaxxx
\left(\frac{\cJ_1}{\ellK(t)-\ellE(t)}\right)^2-\left(\frac{\cJ_2}{(1-t)\ellK(t)-\ellE(t)}\right)^2=\frac{4}{\pi^2 t} \, .
\eexxxa
In fact, for the case of two non-zero spins the evolution equations
of the Neumann model coincide with those of a plane pendulum in
a gravitational field.
Its motion has two phases, either oscillatory or
rotating, which in the present stringy
context translate into the folded or circular type solutions, respectively.

\subsection{B\"acklund transformation and the conserved charges}
One possible way to construct an infinite family of {\it local commuting}
integrals of motion for a sigma model is to use the
B\"acklund transformations. These
transform a solution of the evolution equations into a new one.
The transformation can be usually
constructed by perturbing (``dressing'')
some trial solution.\footnote{The B\"acklund transformations should
not be confused with the so-called dressing transformations whose generators
are subject to a nonabelian algebra (see e.g. \cite{BB}).}

Recall that in light-cone coordinates $\xi,\eta$, where
$\tau=\xi+\eta$, and $\sigma=\xi-\eta$, the evolution equations
for the O(6) sigma model read as \beaxxx \label{ee}
X_{\xi\eta}+(X_\xi \cdot X_{\eta})X=0\, , ~~~~(X\cdot X)=1\, ,\eexxxa
where $X_{\xi}$ denotes the derivative of $X$ w.r.t.~to the
variable $\xi$, and so on, and $(X_\xi \cdot X_{\eta})$ is the scalar
product of two vectors.

Our basic idea is to use the folded
or circular strings as trial solutions and to dress them.
When the new solution is found it can be used
to obtain the values of the local commuting charges
on the trial solution.

In terms of the embedding coordinates $X_i$ the B\"acklund
transformations for the O(6) model can be described by the following
set of equations (see e.g.~\cite{Ogielski:hv})
\beaxxx \nonumber
2\gamma^2 (X(\gamma)+X)_{\xi}&=&(1+\gamma^2)(X(\gamma)\cdot X_{\xi})(X(\gamma)-X)\, ,\\
\label{BT} 2(X(\gamma)-X)_{\eta}&=&-(1+\gamma^2)(X(\gamma)\cdot
X_{\eta})(X(\gamma)+X)\, \eexxxa together with the normalization
conditions \beaxxx \label{norm}
X(0)=X\, ,~~~~~ (X(\gamma)\cdot X(\gamma))=1 \, , ~~~~~
(X(\gamma)\cdot X)=\frac{1-\gamma^2}{1+\gamma^2}\, .
\eexxxa
Here $\gamma$ is a spectral parameter and
$X$ is a trial solution. Solving eqs.(\ref{BT}) we obtain another
solution $X(\gamma)$ of the evolution equations (\ref{ee}) which
is the dressing of our trial solution $X$. The equations (\ref{BT})
admit a solution in terms of a power series in the spectral parameter
$\gamma$
\beaxxx
X(\gamma)=\sum_{k=0}^{\infty}X^{(k)}\gamma^k\, ,
\label{as}
\eexxxa
where the coefficients $X^{(k)}$ can be found recurrently to any desired
order.

The generating function of the local commuting charges of the O(6)
sigma model
\beaxxx
\label{gf}
\cE(\gamma)=\sum_{k=2}^{\infty}\cE_k\gamma^k
\eexxxa
can be
obtained from the dressed solution:
\beaxxx
\label{gfds}
\cE(\gamma)= \int \frac{d\sigma}{4\pi}\Big[ \gamma(X(\gamma)\cdot
X_{\xi})+\gamma^3 (X(\gamma)\cdot X_{\eta}) \Big] \, .
\eexxxa

Let us now discuss the perturbative
construction of the dressed solution $X(\gamma)$.
Plugging the ansatz (\ref{as}) into eqs.(\ref{BT}), (\ref{norm})
it is easy to determine the first few coefficients $X^{(k)}$:
\beaxxx
X^{(0)}=X\, , ~~~X^{(1)}=\frac{2X_{\xi}}{||X_{\xi}||}\, , ~~~
X^{(2)}=\frac{X^{(1)}_{\xi}}{||X_{\xi}||}\, ,
\eexxxa
where we have introduced $||X_{\xi}||=\sqrt{(X_{\xi}\cdot X_{\xi})}$.
The general coefficient $X^{(k)}$ can be obtained recurrently
from $X^{(l)}$ with $l<k$:
\beaxxx X^{(k+1)}=\frac{1}{2||X_{\xi}||}\Big[
Y^{(k)}-\frac{X_{\xi}}{2||X_{\xi}||^2}(Y^{(k)}\cdot X_{\xi}) \Big]
\, , \eexxxa where
 \beaxxx Y^{(k)}=2X^{(k)}_{\xi}-\sum_{l=1}^{k-1}\Big[
(X^{(l)}\cdot X_{\xi})X^{(k-l)}+(X^{(l+1)}\cdot X_{\xi})X^{(k+1-l)} \Big]\, .
\eexxxa

The coefficients $\cE_k$ of the generating function
$\cE(\gamma)$, which are  mutually commuting conserved charges,
are then given  by
\beaxxx
\label{bsf}
\cE_k=\int \frac{d\sigma}{4\pi}\Big[
(X^{(k-1)}\cdot X_{\xi})+ (X^{(k-3)}\cdot X_{\eta}) \Big]\, ,
~~~~k\geq 2 . \eexxxa Here $(X^{(k)}\cdot X_-)=0$ for $k$
negative.


\subsection{Perturbative matching of string and gauge theory charges}
Here we compute the values of the first few conserved charges (\ref{bsf})
on the solution of the Neumann system corresponding to the
folded string and show that up to some
natural linear redefinition they match precisely those found in
one-loop gauge theory. Matching of the charges on the circular type solution
is discussed in \mbox{appendix A.}

The charge $\cE_2$  follows easily upon substituting
the Neumann ansatz into (\ref{bsf}):
\beaxxx
\cE_{2}=\int \frac{d\sigma}{4\pi}(X^{(1)}\cdot X_{\xi})
\, =\int \frac{d\sigma}{2\pi} ||X_{\xi}||
\, =\sqrt{2H}\, .
\eexxxa
Here (the prime denotes a derivative w.r.t.~$\sigma$) 
\beaxxx
\label{ham}
H=\frac{1}{2}\sum_{i=1}^3~{x_i'}^2+w_i^2x_i^2
\eexxxa
is nothing but the Hamiltonian of the Neumann system.
Due to the Virasoro constraint the charge $\cE_2$
coincides with the space-time energy $\cE_2\equiv {\cal E}$
of the string. Note that $H$ is in fact a ``doubly'' conserved quantity:
It is independent with respect to both $\tau$ and $\sigma$.

The next odd integral $\cE_3$ is found to be zero as a consequence of the
relation $(X^{(1)}_{\xi}\cdot X_{\xi})=0$.

In principle, the higher charges, e.g.~the fourth one,
\beaxxx
\cE_4=\int \frac{d\sigma}{4\pi}
\Big[ (X^{(3)}\cdot X_{\xi})+ (X^{(1)}\cdot X_{\eta}) \Big]\, ,
\eexxxa
when computed on the generic Neumann solution $x_i(\sigma)$,
can also be expressed as functionals of $x_i$ and its
derivative\footnote{The higher
derivatives, $x_i''$, $x_i''',\ldots$ are expressed via  $x_i$ and
$x_i''$ by using the equations of motion.}
$x_i'$, i.e. $\cE_n=\cE_n(x_i,x_i')$. However, already for
$\cE_4$ the resulting answer is rather complicated and not very
instructive. Hence, we restrict ourselves to finding
the value of the charges on the special folded string
solution described in the previous section.

On the folded string solution (\ref{fs})
the result of the evaluation of the integral $\cE_4$ can be compactly
written in the form
\beaxxx
\label{I4f}
\cE_4=-\frac{16}{\pi^2 \cE_2}\ellK(t)\Big[
\ellE(t)+(t-1)\ellK(t)\Big]+\frac{32}{\pi^4\cE_2^3}t(t-1)\ellK(t)^4 \, .
\eexxxa
Some comments are in order. The initial result for $\cE_4$ depends on
the Neumann frequencies $w_1$ and $w_2$. We find it convenient
to replace the frequencies by the Jacobi modulus $t$
and the second charge $\cE_2$. To this end we recall \cite{ft4,AFRT}
that for the folded string the periodicity
condition  and the charge $\cE_2$ written in terms of the $w$'s are 
(cf.~eq.(\ref{cspcf}) and eq.(\ref{ham}))
\beaxxx
\label{we}
\ellK(t)^2=\frac{\pi^2}{4}w_{21}^2\, , ~~~~~~
\cE_2^2=w_1^2+tw_{21}^2 \, .
\eexxxa
These formulae allow one to eliminate the $w$'s. We represent the final result
as an expansion in inverse powers of $\cE_2$.

Coming to the next charge $\cE_5$ it can be shown that its integrand is
a total derivative in the variable $\sigma$
and, therefore, $\cE_5$ vanishes.
Evaluation of $\cE_6$ is similar to $\cE_4$, the result is
\beaxxx
\nonumber
\cE_6&=&\frac{2^5}{\pi^2 \cE_2}\ellK(t)\Big[
\ellE(t)+(t-1)\ellK(t)\Big]\\
\nonumber
&-&\frac{2^6}{3\pi^4\cE_2^3}\ellK(t)^3\Big[
(8t-4)\ellE(t)+(t-1)(15t-4)\ellK(t)
\Big] \\
&+&\frac{2^{9}}{\pi^6\cE_2^5}t(t-1)\ellK(t)^5\Big[
\ellE(t)+(3t-2)\ellK(t)
\Big]
-\frac{5\cdot 2^{9}}{\pi^8\cE_2^7}t^2(t-1)^2\ellK(t)^8\, .
\eexxxa

The expressions for $\cE_4$ and $\cE_6$ already suggest
a general pattern: All the odd charges are
zero while the even charges
admit an expansion in odd powers of the inverse second charge.

The conserved charges $\cE_n$ comprise an all-loop result.
To compare with gauge theory we need to extract the
one-loop contribution. Assuming $\cJ=\cJ_1+\cJ_2$ large one first solves
the equation for the Jacobi modulus (\ref{Jmf})
by power series expansion $t=t_0+t_2/\cJ^2+...$.
This implies in particular the following relation between the filling 
fraction $\alpha$ and the modulus $t_0$:
\beaxxx
\label{eqt0}
\alpha=\frac{\cJ_2}{\cJ}=1-\frac{\ellE(t_0)}{\ellK(t_0)}\, .
\eexxxa
Then one can easily verify that the non-trivial string charges admit
the following large $\cJ$ expansion:
\beaxxx
\label{expan}
\cE_n=\delta_{2,n}\cJ+\frac{\epsilon_n^{(1)}}{\cJ}+\frac{\epsilon_n^{(2)}}{\cJ^3}
+\frac{\epsilon_n^{(3)}}{\cJ^5}+...
\eexxxa
Note that only the
second charge scales with $\cJ$ to leading order. In the expansion
(\ref{expan}) all the coefficients $\epsilon_n^{(k)}$ are functions
of $t_0$, while $t_0$ is determined by eq.(\ref{eqt0}). In particular
we have
\beaxxx
\epsilon_2^{(1)}= \frac{2}{\pi^2}\ellK(t_0)
\Big[\ellE(t_0)-(1-t_0)\ellK(t_0)\Big]\, .
\eexxxa
Working out the expansion (\ref{expan}) for $\cE_4$ we find that
the expression for $\epsilon_4^{(1)}$
coincides up to a numerical coefficient
with $\epsilon_2^{(1)}$: $\epsilon_4^{(1)}=-8\epsilon_2^{(1)}$.
Thus, at order $1/\cJ$ the charges $\cE_2$ and $\cE_4$ are not
independent. This suggests that the following redefinition of the
fourth charge should be made
\beaxxx
\label{im4}
\cE_4\to \cE_4+8(\cE_2-{\cal J})\, .
\eexxxa
Indeed, the expansion of the redefined charge starts from the
$1/\cJ^3$ term and we obtain
\beaxxx
\label{cQ4}
{\cal Q}_4\equiv\epsilon_4^{(2)}+8\epsilon_2^{(2)}
=\frac{16}{\pi^4}\ellK(t_0)^3\Big[(2t_0-1)\ellE(t_0)
+(1-4t_0+3t_0^2)\ellK(t_0) \Big] \, .
\eexxxa
Analogously we find that at orders $1/\cJ$ and $1/\cJ^3$
the charge $\cE_6$ is a  linear combination
of the lower charges $\cE_2$ and $\cE_4$. Again this redundant
information contained in $\cE_6$ can be removed by redefining it as
\beaxxx
\label{im6}
\cE_6\to \cE_6+\frac{128}{3}(\cE_2-\cJ)+\frac{22}{3}\cE_4\, .
\eexxxa
The expansion of the redefined charge starts from $1/\cJ^5$
and the coefficient reads
\beaxxx
\label{cQ6}
&&{\cal Q}_6\equiv\epsilon_6^{(3)}
+\frac{128}{3}\epsilon_2^{(3)}+\frac{22}{3}\epsilon_4^{(3)}
=\\
\nonumber
&&~~~~~~~~~
\frac{2^{9}}{3\pi^6}\ellK(t_0)^5\Big[(1-6t_0+6t_0^2)\ellE(t_0)
+(t_0-1)(1-8t_0+10t_0^2)\ellK(t_0)\Big] \, .\eexxxa
Thus, we see that after redefinition the (improved) charges
scale homogeneously with inverse powers of $1/\cJ$, i.e. 
like ${\cal Q}_{n}/\cJ^{n-1}
 $,
as was the case for the conserved charges of the one-loop gauge theory.

Concerning the next two charges we just point out that they 
require the following redefinition
\beaxxx
\nonumber
\cE_8&\to & \cE_8+\frac{1024}{5}(\cE_2-\cJ)+\frac{201}{5}\cE_4+\frac{44}{5}\cE_6 \, , \\
\nonumber
\cE_{10}&\to &\cE_{10}+\frac{32768}{35}(\cE_2-\cJ)+\frac{6922}{35}\cE_4+\frac{1898}{35}\cE_6+\frac{74}{7}\cE_8 \,
\eexxxa
and the corresponding values of the improved charges are
{\small
\beaxxx
\nonumber
{\cal Q}_8&=&\frac{2^{11}}{\pi^8}\ellK(t_0)^7
\Big[(2t_0-1)(1-10t_0+10t_0^2)\ellE(t_0) \\
\nonumber
&&~~~~~~~~~~~~~~~~~~~~~+(t_0-1)(-1+15t_0-45t_0^2+35t_0^3)\ellK(t_0)
\Big]\, , \\
\nonumber
{\cal Q}_{10}&=&\frac{2^{17}}{5\pi^{10}}\ellK(t_0)^9
\Big[(1-20t_0+90t_0^2-140t_0^3+70t_0^4)\ellE(t_0) \\
\nonumber
&&~~~~~~~~~~~~~~~~~~~~~+(t_0-1)(1-24t_0+126t_0^2-224t_0^3+126t_0^4)\ellK(t_0)
\Big]\, .
\eexxxa
}

Now we will demonstrate that the
improved large ${\cal J}$ string theory charges ${\cal Q}_{2 n}$
and the one-loop gauge theory charges
$Q_{2 n}$ are actually the same. The charges ${\cal Q}_{2 n}$ depend
on the string modulus $t_0$, while the $Q_{2 n}$ are governed by $q$.
Matching ${\cal Q}_2\equiv \epsilon_2^{(1)}$ with $Q_2$ requires that
the string modulus is related to the gauge theory
modulus by a Gauss-Landen transformation \cite{AFRT,BFST}, which for the
folded string solution is of the form
\beaxxx
\label{GL}
t_0=-\frac{(1-\sqrt{1-q})^2}{4\sqrt{1-q}}\, .
\eexxxa
Under this transformation one has
\bexxx\label{eq:transform1}
\ellK(t_0)=(1-q)^{1/4}\ellK(q),\quad\ \
\ellE(t_0)=\frac{1}{2} (1-q)^{-1/4} \ellE(q)+\frac{1}{2}
(1-q)^{1/4}\ellK(q) \ .
\eexxx
Remarkably this maps not only the energy, but {\it all} higher
large ${\cal J}$ string charges ${\cal Q}_{2 n}$ onto the one-loop
gauge charges $Q_{2 n}$. Indeed, transforming the first few
({\it cf} eqs.(\ref{cQ4}), (\ref{cQ6}))
${\cal Q}_{2 n}$ with (\ref{GL}), (\ref{eq:transform1}) we obtain
\beaxxx
\label{foldedmatch}
{\cal Q}_2&=&-\frac{1}{4\pi^2}Q_2\, , ~~~
{\cal Q}_4=\frac{3}{8\pi^4}Q_4\, , ~~~
{\cal Q}_6=-\frac{5}{12\pi^6}Q_6\, , \\
\nonumber
{\cal Q}_8&=&\frac{7}{16\pi^8}Q_8\, ,~~~
{\cal Q}_{10}=-\frac{9}{20\pi^{10}}Q_{10}\, , \quad \ldots \quad .
\eexxxa
Up to unessential numerical prefactors the improved
``one-loop'' string charges ${\cal Q}_n$ coincide
with the gauge theory charges $Q_n$, {\it cf} eqs.(\ref{eq:Q4})
in section 2.2. An all orders proof will be presented in chapter 4,
and the exact matching formula turns out to be
\beaxxx
\label{match1}
{\cal Q}_{2k}=(-1)^k\frac{2k-1}{4k\pi^{2k}}Q_{2k}\, .
\eexxxa

We note a certain similarity of the higher terms $\epsilon_2^{(n)}$
in the expansion of $\cE_2$, corresponding to the currently unknown
eigenvalues of the dilatation operator at higher loops,
and the ``one-loop'' higher charges ${\cal Q}_n$. For instance,
\beaxxx
\epsilon_2^{(2)}=
-\frac{2}{\pi^4}\ellK(t_0)^3\Big[(2t_0-1)\ellE(t_0)
+(1-t_0)^2\ellK(t_0) \Big] \,
\eexxxa
is similar to but different from ${\cal Q}_4$ in eq.(\ref{cQ4})
due to the additional term $\epsilon_4^{(2)}$. The difference is of course
expected, otherwise
the higher loop dilatation operator would commute with a higher one-loop
charge, {\it i.e.}, integrability could be extended to higher loops
without any deformation, which is certainly not the case.

With the dressed solution $X(\gamma)$ at hand we may ask
for the corresponding values of the commuting charges,
as we did for the trial solution $X$.
Performing a perturbative (in $\gamma$) computation of the spins
$\cJ_i$ and the first few charges $\cE_n$ on $X(\gamma)$,
we find that neither the spins nor the charges depend on the deformation
parameter $\gamma$; higher corrections $Z^{(k)}$ to the trial solution
simply do not contribute. 
Another important point is that $X(\gamma)$ respects the Virasoro
constraints.
All this shows, in fact, that the
dressed and the trial solutions
differ in a very mild fashion. In the next section we will use
this observation to solve the B\"acklund equations exactly and to find
all string commuting charges.

\subsection{Exact generating function of commuting charges in string theory}

Having gained some experience from perturbative calculus
we now want to determine an exact (to all orders in $\gamma$)
generating function for the commuting charges in string theory.
A direct approach to sum the series (\ref{as})
seems hopeless
due to increasing complexity of the coefficients $X^{(k)}$.
Therefore we have to better understand the meaning
of the B\"acklund transformations, and reformulate
the problem in a suitable way.

A first observation is that
the B\"acklund transformation ``almost''
preserves the Neumann form
of the trial solution. An easy way to see this is to write
the perturbative solution in terms of the
complexified embedding coordinates $Z_i=X_{2i-1}+iX_{2i}$, $i=1,2,3$,
where $X_1,\ldots ,X_6$ are the original real variables of the
O(6) sigma model. Then
the first few coefficients of the dressed solution are (\ref{as})
\beaxxx
\nonumber
&&Z_i=x_i~e^{iw_i\tau}\, , \\
&&Z_i^{(1)}=
\frac{2}{\sqrt{\cE_2}}(\partial_{\sigma}+iw_i)x_i~e^{iw_i\tau} \, , \\
\nonumber
&&Z_i^{(2)}=
\frac{2}{\cE_2}(\partial_{\sigma}+iw_i)^2x_i~e^{iw_i\tau} \,
\eexxxa
and so on (the higher terms look more complicated).
To get these formulae one uses the fact that
$\cE_2$ is an integral for both the sigma model
($\tau$-independent) and for the Neumann system
($\sigma$-independent). Due to the phase exponents,
$e^{iw_i\tau}$,
the deformed solution is still of rotating type,
but the $\sigma$-dependent coefficients become complex,
in contrast to the real variables $x_i$ of the trial solution.
Thus, the B\"acklund transformation allows one
to construct more general solutions of the O(6) model which
correspond to the ``complexified''
Neumann ansatz:
\beaxxx
\label{BA}
Z_i=r_i(\sigma,\gamma)~e^{i\alpha_i(\sigma, \gamma)}~e^{iw_i\tau} \, ,
\eexxxa
where, in addition to the radial coordinates $r_i$,
the phase shifts $\alpha_i$ come into play.

Another important observation comes from the analysis of the phase shifts
$\alpha_i$ for the first few perturbative coefficients of the trial solution:
The phases $\alpha_i$ do not depend on $\sigma$, {\it i.e.}~they
are functions of the spectral parameter only.
This fact implies that the variables
$r_i(\sigma,\lambda)$ solve the evolution equations of the Neumann system:
\beaxxx
r_i''=-w_i^2r_i-r_i\sum_{j=1}^3(r_j'^2-w_j^2r_j^2)\, .
\eexxxa
Thus, $r_i(\sigma,\gamma)$ is a one-parameter family of solutions
obeying the initial condition $r_i(\sigma,0)=x_i(\sigma)$.
On the other hand we already know {\it all} two-spin solutions\footnote{The B\"acklund transformation preserves all the global charges of the O(6) model.} of the
Neumann system
corresponding to the folded and the circular strings \cite{AFRT}.
They indeed form a one-parameter family, where the parameter
is related to rigid shifts of the world-sheet variable $\sigma$:
$\sigma\to \sigma+\mbox{const}$. Therefore we can
immediately write down the generic folded string solution depending on the
spectral parameter:
\beaxxx
\label{BS}
r_1(\gamma)=\dn(\mu+\nu,t)\, , ~~~~~~
r_2(\gamma)=\sqrt{t}~\sn(\mu+\nu,t)\, ,~~~~~r_3(\gamma)=0 \, ,
\eexxxa
where we introduced the concise notation
$\mu=\sigma \sqrt{w_{21}^2}$. The variable $\nu$ is a yet unknown
function of the spectral parameter $\nu\equiv \nu(\gamma)$; however we
know that it cannot
depend on $\sigma$. Requiring the solution to be periodic we obtain the
usual periodicity condition for the folded string which we repeat here
for convenience:
$
\frac{\pi}{2}\sqrt{w_{21}^2}=\ellK(t)\,
$.

Thus we considerably constrained the ansatz for the B\"acklund solution:
The only unknowns left are the two phases $\alpha_1$,
$\alpha_2$ and the shift function $\nu$, all of which depend solely on the
spectral parameter $\gamma$. Now we can use the B\"acklund equations (\ref{BT})
to determine $\alpha_1,\alpha_2,\gamma$, and furthermore check the consistency
of our ansatz. The corresponding solution is found
in appendix A and we use it to obtain the exact generating function
for the commuting string charges. The final result reads
\beaxxx
\label{main}
{\cal E}(\gamma)=\frac{4\gamma^3}{\pi(1+\gamma^2)}
\frac{\sqrt{(1-z)(1-tz)}}{\sqrt{z}}~\Pi(tz,t)\, ,
\eexxxa
where the function $z=\sn^2\nu$ satisfies
\beaxxx
\label{aux}
1-\frac{w_1^2}{w_{21}^2}\frac{z}{1-z}-\left(\frac{1-\gamma^2}{1+\gamma^2}\right)^2\frac{1}{1-tz}=0\, .
\eexxxa
The last equation implies that $z$ can be expanded
in even powers of $\gamma$, starting at ${\cal O}(\gamma^2)$.
Carefully developing the
$\gamma$-expansion of the elliptic integral $\Pi$ and eliminating the $w$'s
with the help of eqs.(\ref{we})
\beaxxx
\label{wf}
w_1=\frac{1}{\pi}\sqrt{\pi^2\cE_2^2-4t\ellK(t)^2}\, ,~~~~~~
w_2=\frac{1}{\pi}\sqrt{\pi^2\cE_2^2-4(t-1)\ellK(t)^2} \, ,
\eexxxa
one can check
that the formula (\ref{main}) correctly reproduces our previous
findings for the
first few non-trivial charges $\cE_2$, $\cE_4$, $\cE_6$ and so on.  It also
shows that all odd charges are zero.

To complete the solution we recall that the energy $\cE_2$ and the modulus $t$
are functions of the spins determined by the following two equations:
\beaxxx
\label{epcf}
\left(\frac{\cE_2}{\ellK(t)}\right)^2-
\left(\frac{{\cal J}_1}{\ellE(t)}\right)^2
=\frac{4}{\pi^2}t \, , ~~~~
\left(\frac{\cJ_2}{\ellK(t)-\ellE(t)}\right)^2-\left(\frac{\cJ_1}{\ellE(t)}\right)^2=\frac{4}{\pi^2} \, .
\eexxxa
Thus, all string charges are implicit functions of the spins $\cJ_i$.

To better appreciate the result obtained let us note
that the function $\cE(\gamma)$
depends explicitly on the energy $\cE_2$.
Given the energy $\cE_2$ and the modulus $t$, all the higher charges
are uniquely determined by (\ref{main}). From the point of view
of the Neumann integrability this result is expected:
the Neumann system has two integrals of motion which are essentially
$\cE_2$ and $t$. These integrals completely constrain the solution
(up to a constant shift in $\sigma$), and we certainly do not need to
use the higher charges in order to derive the energy. All higher
charges of the sigma model embedding are then determined
{\it a posteriori}.
It is rather intriguing that this structure appears to
differ significantly, on a conceptual level,
from the one found on the gauge theory side. There all charges,
generated by the transfer matrix operator,
are required in order to apply the Bethe ansatz and obtain 
the solution of the spin chain energy spectrum. 
The final results, when restricted to one-loop, nevertheless
match precisely.

Finally let us also note that eqs.(\ref{wf}), (\ref{epcf}) together with eq.(\ref{eqt0}) allows one to rewrite eq.(\ref{aux}) in the form
\beaxxx
\label{anff}
\left(\frac{1-\gamma^2}{1+\gamma^2}\right)^2=
\frac{1-t z}{1- z} \left[1-
\left( \frac{\ellK(t_0)-\ellE(t_0)}
{\ellK(t)-\ellE(t)}  \right)^2
\frac{\pi^2}{4} \frac{\cJ^2}{\ellK(t_0)^2}~z
\right]
\eexxxa
This formula is most suitable  for generating
$\gamma$- and $\cJ$-expansions of $z= z(\gamma,\cJ)$
and, as the consequence, of $\cE(\gamma)$. Indeed, the second formula in 
(\ref{epcf}) and eq.(\ref{eqt0}) can be used to determine
\beaxxx
\label{expm}
t=t_0+\frac{t_2}{\cJ^2}+\frac{t_4}{\cJ^4}+...\, ,
\eexxxa
where all the coefficients $t_{2k}$, $k>0$ are functions of $t_0$.
Then all the coefficients in the expansion of $z$ appear to depend on $t_0$.

The derivation of the generating function for commuting charges related to
the circular string is very similar. The exact solution of the corresponding
B\"acklund equations is found in appendix A. Remarkably,
the generating function
for the circular string turns out to be of the identical form (\ref{main}).
The only difference lies in the equation for the shift function $\nu$:
\beaxxx 
\label{auxc}
\frac{tw_1^2}{w_{12}^2}z +
\left(\frac{1-\gamma^2}{1+\gamma^2}\right)^2\frac{1-tz}{1-z}-1=0\,
, ~~~z=\sn^2\nu \, .
 \eexxxa
This formula is supplemented by the expressions for the $w$'s via the energy
and the modulus
 \beaxxx \nonumber w_1=\frac{1}{\pi}\sqrt{\pi^2{\cal
E}_2^2+4(t-1)\ellK(t)^2}\, ,
~~~~~w_2=\frac{1}{\pi}\sqrt{\pi^2{\cal E}_2^2-4\ellK(t)^2}\, \eexxxa
and by two transcendental equations yielding the energy
and the modulus as functions of the spins
\beaxxx
\label{eq:circularenergy}
&&\left(\frac{\cE_2}{\ellK(t)}\right)^2-
\left(\frac{t{\cal J}_1}{\ellK(t)-\ellE(t)}\right)^2
=\frac{4}{\pi^2}(1-t) \, , \\
\nonumber
&&\left(\frac{\cJ_1}{\ellK(t)-\ellE(t)}\right)^2-\left(\frac{\cJ_2}{(1-t)\ellK(t)-\ellE(t)}\right)^2=\frac{4}{\pi^2 t} \, .
\eexxxa
These expressions together with
the formula relating the modulus $t_0$ with the filling fraction
$\alpha$ for the circular string:
\beaxxx
\alpha=\frac{\cJ_2}{\cJ}\, , ~~~~\alpha=1-\frac{1}{t_0}\left(1-\frac{\ellE(t_0)}{\ellK(t_0)}  \right)
\eexxxa
allow one to cast the expression (\ref{auxc}) in the form
\beaxxx
\label{ancc}
\left(\frac{1-\gamma^2}{1+\gamma^2}\right)^2=
\frac{1-z}{1-t z} \left[ 1-\frac{t^2}{t_0^2}
\left( \frac{\ellK(t_0)-\ellE(t_0)}
{\ellK(t)-\ellE(t)}  \right)^2
\frac{\pi^2}{4} \frac{\cJ^2}{\ellK(t_0)^2}~z \right].
\eexxxa
The explicit ``perturbative'' matching of the first few charges generated by
$\cE(\gamma)$ with their gauge theory counterparts is discussed
in appendix A. 

\section{Matching the entire infinite tower of commuting charges}

We so far derived the exact generating functions of one-loop gauge
charges, eq.(\ref{eq:resolvent2r}) and eq.(\ref{eq:resolvent2}), as well
as the exact generating functions of (unimproved) string charges,
eqs.(\ref{main}),(\ref{epcf}),(\ref{anff}) and 
eqs.(\ref{main}),(\ref{eq:circularenergy}),(\ref{ancc}),
and we showed that the first few charges indeed coincide if we
(a) linearly redefine (improve) the string charges, (b) restrict
to one-loop and (c) apply a Gauss-Landen transformation.
In this final chapter we will apply this three-step procedure to the
entire infinite tower of string charges. 
In other words, we will {\it derive} the infinite number of 
gauge charges from string theory. To be specific we will treat 
the case of the folded string first, and work out the circular
case in appendix A.2.

The idea is to implement the freedom of linear redefinition on
a functional level. In particular, we are clearly allowed to replace
$\gamma^2$ by a new parameter $\mu^2$ which
can be an arbitrary formal, even power series in $\gamma^2$ such that
$\mu^2=\gamma^2+{\cal O}(\gamma^4)$. Ideally we would like to find
an improved spectral parameter $\mu$ such that the transformed
string resolvent eq.(\ref{main}) only generates terms in the combination
$\mu^{2 k} {\cal J}^{-2k+1} +{\cal O}(\mu^{2 k} {\cal J}^{-2k-1})$.
The coefficient multiplying $\mu^{2 k}$ would then correspond to the
fully improved string charge. Inspection of eq.(\ref{anff}) shows
that this can be almost (see below) achieved by defining $\mu$ through
\beaxxx
\label{imprsp}
1-4\mu^2=\left(\frac{1-\gamma^2}{1+\gamma^2}\right)^2 \, .
\eexxxa
Eq.(\ref{anff}) becomes
\beaxxx
\label{modz}
\frac{\mu^2}{\pi^2\cJ^2}=\frac{tz}{4\pi^2\cJ^2}
+\left(\frac{\ellE(t_0)}{4\ellK(t_0)\ellE(t)}\right)^2\frac{z(1-tz)}{1-z}
\, , 
\eexxxa
and we easily verify that the auxiliary parameter $z$ now indeed 
expands in the improved fashion. Turning our attention to
eq.(\ref{main}), we see that we should drop
an overall function in $\mu$ which expands like $1+{\cal O}(\mu^2)$,
as it would
again be a source of unwanted, low powers in $1/{\cal J}$ at
${\cal O}(\mu^{2 k})$ (this is allowed since it once more corresponds
to a linear redefinition). This way we find the following 
improved generating function
\beaxxx
\label{mainren}
{\cal E}(\gamma)\to \tilde{{\cal E}}(\mu)=\frac{4 \mu^3}{\pi}
\frac{\sqrt{(1-z)(1-tz)}}{\sqrt{z}}~\Pi(tz,t)\, .
\eexxxa
which generates linearly redefined string charges
\beaxxx
\label{gfmod}
\tilde{{\cal E}}(\mu)=\sum_{k=1}^{\infty}\tilde{\cE}_{2 k}\mu^{2 k}
\eexxxa
However, in the line preceding eq.(\ref{imprsp}) we wrote ``almost''
since, unfortunately, 
due to the overall power $\mu^3$ in eq.(\ref{mainren}),
the redefined charges $\tilde{\cE}_{2 k}$ now expand as 
$\sim \mu^{2 k} {\cal J}^{-2k+3} +{\cal O}(\mu^{2 k} {\cal J}^{-2k+1})$.
This is much better as the old resolvent eq.(\ref{main}), but
in principle one further linear redefinition of the charges 
$\tilde{\cE}_{2 k} \rightarrow \bar{\cE}_{2 k}$
is needed\footnote{Our derivation only requires the observation that
$\bar{\cE}_{2 k}=\tilde{\cE}_{2 k}+c_{k-1}~\bar{\cE}_{2 k-2}$,
which is the general form of the linear transformation mapping the
charges $\tilde{\cE}_{2 k}$ to the fully-improved charges 
$\bar{\cE}_{2 k}$.
Using different arguments one can show that the a priori unknown
constants $c_k$ are given by $c_k=\frac{8k}{2k-1}$.
}:
\beaxxx
\label{eq:improv}
\bar{\cE}_{2 k}=\tilde{\cE}_{2 k}+4~\frac{2k-2}{2k-3}~\bar{\cE}_{2 k-2}.
\eexxxa
The $\bar{\cE}_{2 k}$ are the properly scaling, fully improved
string charges, requiring a further functional transform 
$\tilde{\cE}(\mu) \rightarrow \bar{\cE}(\mu)$.
However, we see from eq.(\ref{eq:improv}) that to {\it leading}
order in $1/{\cal J}$, the ``nearly-improved'' charges $\tilde{\cE}_{2 k}$
are proportional to the ``fully-improved'' charges $\bar{\cE}_{2 k-2}$!
We can therefore introduce a large-${\cal J}$ 
generating function 
\beaxxx
\label{eq:stringoneloop}
{\cal Q}(u)=\lim_{{\cal J} \rightarrow \infty} 
\frac{1}{2}~\frac{1-\mu^{-2}~\tilde{\cE}(\mu)}{\cJ}
\qquad {\rm with} \qquad
u^2=-\frac{\mu^2}{\pi^2\cJ^2},
\eexxxa
where we deliberately use the same notation 
$u$ as in chapter 2 since it will become clear in a moment 
that $u$ can indeed be identified with the spectral parameter
of the gauge theory resolvent. Restriction to
``one-loop'' corresponds to keeping $u$ finite and
sending $\cJ$ to infinity.  
From eq.(\ref{mainren}) the generating function 
in (\ref{eq:stringoneloop}) becomes
\beaxxx
\label{grbeforeGL}
{\cal Q}(u)=\frac{1}{2}\Big[
1-4u\frac{\sqrt{(1-z)(t_0z-1)}}{\sqrt{z}}\Pi(t_0z,t_0)\,
\Big]. 
\eexxxa
Furthermore, in this limit eq.(\ref{modz})
simplifies drastically and we get 
\beaxxx 
\label{modz1}
u^2+\frac{1}{16\ellK(t_0)^2}\frac{z(1-t_0z)}{1-z}=0
\, . \eexxxa 
We pick up the root which in the vicinity of $u\sim 0$ behaves as $z\sim u^2$, namely 
\beaxxx 
\label{varz}
z=
\frac{1-16\ellK(t_0)^2u^2}{2t_0}\left(1
-\sqrt{
1+\frac{64t_0\ellK(t_0)^2u^2}{\Big(1-16\ellK(t_0)^2u^2\Big)^2}}\right)  \, .
 \eexxxa 
The final step consists in applying the Gauss-Landen transformation 
(\ref{GL}) to eq.(\ref{grbeforeGL}). In particular, one can prove
that the following modular transformation formula is valid
\beaxxx 
\Pi(t_0z,t_0)=\frac{2\sqrt{ab}}{4ab+(a-b)^2v}\left[\frac{1}{4}+\frac{a^2}{b}\sqrt{\frac{b^2-u^2}{a^2-u^2}}\Pi\Big(
-q\frac{u^2}{a^2-u^2},q\Big)\right]\, .
\eexxxa
Here $t_0$ is related to $q$ by eq.(\ref{GL}), the variables $a$ and $b$
are given by eqs.(\ref{eq:ab}), and $v$ is an image of the variable $z$
in eq.(\ref{varz}) under the Gauss-Landen transformation:
\beaxxx
v=2\frac{u^2-ab}{(a-b)^2}\left(1-\sqrt{1-\frac{(a-b)^2u^2}{(u^2-ab)^2}}~  \right) \, .
\eexxxa
With these formulae at hand we find that after application of 
the Gauss-Landen transformation 
the large ${\cal J}$  folded string spectral curve ${\cal Q}(u)$, 
eq.(\ref{grbeforeGL}), becomes 
\beaxxx
{\cal Q}(u)=H(u)+\frac{\alpha}{2}.
\eexxxa
where $H(u)$ is {\it precisely} the one-loop folded gauge theory resolvent 
as written in eq.(\ref{eq:resolvent2r}), q.e.d.

\vskip 0.5cm
\noindent
{\bf Note added.}  After this work was completed we learned that the 
relation between the
Bethe resolvent and the properly defined higher gauge theory charges  
was independently derived by J.~Engquist, J.A.~Minahan and 
K.~Zarembo in the interesting paper \cite{EMZ}.

\section*{Acknowledgments}
The work of  G.A. was supported in part
by the European Commission RTN programme HPRN-CT-2000-00131 and
by RFBI grant N02-01-00695. We would like to thank
Niklas Beisert, Sergey Frolov, Jan Plefka, Jorge Russo, Arkady
Tseytlin and Kostya Zarembo for useful discussions.

\appendix

\section{Exact solution of the B\"acklund equations and generating function}
Using the ansatz (\ref{BA}) for the B\"acklund solution we write down the generating function (\ref{gfds})
\beaxxx
\label{gfint}
\cE(\gamma)=\gamma\int_0^{2\pi}\frac{d\sigma}{4\pi}r_i(\gamma)
\Big[ (1-\gamma^2) \cos\alpha_ix_i'+ (1+\gamma^2)w_i\sin\alpha_i x_i)\Big]
\eexxxa
Here $r_i(\gamma)$ is the B\"acklund solution (\ref{BS}) and
$x_i$ is the trial solution: $x_i=r_i(0)$.
As was discussed in the main text, to compute $\cE(\gamma)$
one has to find the phases $\alpha_i(\gamma)$ and the shift function $\nu(\gamma)$.

\subsection{Folded case}
We start with the folded string described by eqs.(\ref{fs});
the spectral dependent solution $r_i(\gamma)$ is given by (\ref{BS}).
Consider the normalization condition (\ref{norm})
\beaxxx
\label{NC}
\cos\alpha_1 r_1(\gamma)x_1+\cos\alpha_2 r_2(\gamma)x_2=\frac{1-\gamma^2}{1+\gamma^2} \, .
\eexxxa
Specifying $\sigma=0$ or $\sigma=\frac{\pi}{2}$ one finds
\beaxxx
\label{cosin}
\cos\alpha_1=\frac{1-\gamma^2}{1+\gamma^2}\frac{1}
{\mbox{dn}\nu }\, \, , ~~~~~~
\cos\alpha_2=\frac{1-\gamma^2}{1+\gamma^2}\frac{\mbox{cn}\nu }
{\mbox{dn}\nu }\, .
\eexxxa
Plugging the found solution back into eq.(\ref{NC})
we obtain the following relation
\beaxxx
\nonumber
\dn(\mu+\nu)\dn\nu+
t\cn \mu~\sn \mu~\sn(\mu+\nu)=\dn\nu \, , ~~~~\mu\equiv \sigma\sqrt{w_{21}^2}
\eexxxa
which
must hold for any value of $\sigma$. This is indeed the case
as one can verify by using the addition formulas for elliptic functions.

It is convenient to separate the real and the
imaginary parts of the B\"acklund equations.
Then one can show that with the choice
(\ref{cosin}) the real part of eqs.(\ref{BT}) is identically satisfied while
the imaginary part implies the relations
\beaxxx
\label{sin}
\sin\alpha_1=\frac{w_1}{\sqrt{w_{21}^2}}\frac{\sn \nu}{\cn \nu} \, , ~~~~~~~
\sin\alpha_2=\frac{w_2}{\sqrt{w_{21}^2}}\sn \nu \, .
\eexxxa
Now
the identity $\cos^2\alpha_1+\sin^2\alpha_1=1$ gives an
equation for the unknown function
$\nu$:
\beaxxx
\label{sin1}
\frac{w_1^2}{w_{21}^2}\frac{\sn^2 \nu}{\cn^2\nu}
+\left(\frac{1-\gamma^2}{1+\gamma^2}\right)^2
\frac{1}{\dn^2 \nu}=1 \,
\eexxxa
which is equivalent to eq.(\ref{aux}).

Finally using eqs.(\ref{cosin}) and (\ref{sin}) one can prove that
the $\sigma$-dependence of the eqs.(\ref{BT}) is identically satisfied
due to elliptic function identities. This shows consistency of our
ansatz for the B\"acklund solution. Thus we have found the exact solution of the
B\"acklund equations.

It is worth stressing that the B\"acklund solution we found 
has the same dual CFT operator as the trial solution. 
Indeed, the phases in eq.(\ref{BA}), which are independent of 
$\tau$ and $\sigma$, 
can be gauged away by the rigid U(1)$\times$U(1)$\times$U(1)
rotation, while the function $\nu(\gamma)$ can be removed by 
the constant shift of $\sigma$ (the leftover reparametrization invariance).
However, the B\"acklund equations themselves are not invariant 
w.r.t. these symmetries and that is why they lead to a 
particular solution with non-trivial
$\gamma$-dependence.

To obtain the the generating function
the following integration formulae are helpful
\beaxxx
\nonumber
\int_0^{2\pi}\frac{d\sigma}{4\pi}r_1(\lambda)r_1(0)&=&
\frac{\dn\nu}{2\ellK(t)\sn^2\nu}\Big[\ellK(t)-\cn^2\nu\Pi(t\sn^2\nu,t)
\Big]\, ,\\
\nonumber
\int_0^{2\pi}\frac{d\sigma}{4\pi}r_2(\lambda)r_2(0)&=&
\frac{\cn\nu~\dn\nu}{2\ellK(t)\sn^2\nu}\Big[\Pi(t\sn^2\nu,t)-\ellK(t) \Big]\, ,\\
\nonumber
\int_0^{2\pi}\frac{d\sigma}{4\pi}r_1(\lambda)r_1'(0)&=&
\frac{\cn\nu}{\pi\sn^3\nu}\Big[\dn^2\nu(\ellK(t)-\Pi(t\sn^2\nu,t))+\sn^2\nu(\ellK(t)-\ellE(t)) \Big]\, ,\\
\nonumber
\int_0^{2\pi}\frac{d\sigma}{4\pi}r_2(\lambda)r_2'(0)&=&
\frac{1}{\pi\sn^3\nu}\Big[\sn^2\nu\ellE(t)-\dn^2\nu\ellK(t)
+\cn^2\nu\dn^2\nu\Pi(t\sn^2\nu,t)\Big]\, .
\eexxxa
They can be easily derived by using the addition formulae for elliptic functions
and the integral representation for the complete elliptic integral of the
third kind $\Pi(u,t)$. Using these formulae we obtain
\beaxxx
\label{main1}
\cE(\gamma)&=&\frac{\gamma}{\pi(1+\gamma^2)}
\left[
(1-\gamma^2)^2\frac{\cn \nu}{\sn \nu \dn \nu}
\Big[\ellK(t)-\dn^2 \nu\Pi(t\sn^2\nu,t) \Big] \right.   \\
\nonumber
&+&\left. (1+\gamma^2)^2\Big[
\frac{\dn \nu \cn \nu}{\sn \nu}\Pi(t\sn^2\nu,t)
-\left(\frac{w_2^2\cn \nu}{w_{21}^2}-
\frac{w_1^2}{w_{21}^2\cn \nu}  \right)\frac{\dn \nu}{\sn \nu}\ellK(t)\Big]
\right] \, ,
\eexxxa
The last step consists in
expressing the quantities $w_1^2/w_{21}^2$ and
$w_2^2/w_{21}^2$ from eq.(\ref{sin1})
\beaxxx
\frac{w_1^2}{w_{21}^2}&=&\frac{\cn^2\nu}{\sn^2\nu}-\left(\frac{1-\gamma^2}{1+\gamma^2}\right)^2\frac{\cn^2\nu}{\sn^2\nu\dn^2\nu}\, ,\\
\frac{w_2^2}{w_{21}^2}&=&\frac{1}{\sn^2\nu}-\left(\frac{1-\gamma^2}{1+\gamma^2}\right)^2\frac{\cn^2\nu}{\sn^2\nu\dn^2\nu}\,
\eexxxa
and substituting them into eq.(\ref{main1}).
In this way we obtained our final result
(\ref{main}).

\subsection{Circular case}

Here we will also treat the circular string (\ref{cs}), i.e.~we will find
its curve and charges on the string side and then match it to the
gauge theory results in section 2.2. The presentation
will be more succinct as the one above.

One starts with the following ansatz for the B\"acklund
solution
\beaxxx
r_1(\gamma)=\sn(\mu+\nu), \quad
r_2(\gamma)=\cn(\mu+\nu), \quad r_3(\gamma)=0
\quad {\rm with} \quad
\mu=\sigma \sqrt{\frac{w_{12}^2}{t}}.
\eexxxa
Then
the following formulae solve the B\"acklund equations for the
circular string:
 \beaxxx \cos\alpha_1&=&
\frac{1-\gamma^2}{1+\gamma^2}\frac{\dn\nu}{\cn\nu}\, ,~~~~
\cos\alpha_2= \frac{1-\gamma^2}{1+\gamma^2}\frac{1}{\cn\nu}\, , \\
\sin\alpha_1&=&\frac{\sqrt{t}~w_1}{\sqrt{w_{12}^2}}\sn\nu \,
,~~~~~
\sin\alpha_2=\frac{\sqrt{t}~w_2}{\sqrt{w_{12}^2}}\frac{\sn\nu}{\dn\nu}
\, .
 \eexxxa
The equation for the shift function $\nu$ is \beaxxx
\label{sfc}
\frac{tw_1^2}{w_{12}^2}\sn^2\nu +
\left(\frac{1-\gamma^2}{1+\gamma^2}\right)^2\frac{\dn^2\nu}{\cn^2\nu}=1\, ,
 \eexxxa
which can be written as \beaxxx \frac{tw_1^2}{w_{12}^2}z +
\left(\frac{1-\gamma^2}{1+\gamma^2}\right)^2\frac{1-tz}{1-z}-1=0\,
, ~~~z=\sn^2\nu \, .
 \eexxxa
The frequencies $w_i$ are found from the equations \cite{AFRT}
\beaxxx
\frac{\pi^2}{4}\frac{w_{12}^2}{t}=\ellK(t)^2\, , ~~~~~ {\cal
E}_2^2=w_2^2+\frac{w_{12}^2}{t} \, , ~~~~~w_1>w_2
 \eexxxa
and they are \beaxxx \nonumber w_1=\frac{1}{\pi}\sqrt{\pi^2{\cal
E}_2^2+4(t-1)\ellK(t)^2}\, ,
~~~~~w_2=\frac{1}{\pi}\sqrt{\pi^2{\cal E}_2^2-4\ellK(t)^2}\, .\eexxxa

The exact generating function is \beaxxx {\cal E}(\gamma) \label{csgf}
&=&
\frac{\gamma}{\pi(1+\gamma^2)}\Big[(1-\gamma^2)^2\frac{\dn\nu}{\sn\nu\cn\nu}\Big[\ellK(t)-\cn^2\nu\Pi(t\sn^2\nu)\Big]\\
\nonumber &&+(1+\gamma^2)^2\Big[
\frac{\cn\nu\dn\nu}{\sn\nu}\Pi(t\sn^2\nu,t)-\left(\frac{w_1^2\dn\nu}{w_{12}^2}-\frac{w_2^2}{w_{12}^2\dn\nu}
\right)\frac{\cn\nu}{\sn\nu} \ellK(t)\Big]\, . \eexxxa
Again,
excluding the $w$'s with the help of eq.(\ref{sfc}) for the shift
function $\nu$ we can simplify the expression (\ref{csgf}) to get
\beaxxx
\label{ccc}
{\cal
E}(\gamma)=\frac{4\gamma^3}{\pi(1+\gamma^2)}\frac{\cn\nu\dn\nu}{\sn\nu}\Pi(t\sn^2\nu
,t)\, . \eexxxa
Quite remarkably, the generation function for the circular string
turns out to be exactly of the same form as as the one of the folded
string. The
difference lies of course in the equation for the shift function.

As in the folded string case all odd charges
vanish for the circular string configuration.
From eq.(\ref{ccc}) it is easy to obtain
the explicit expressions for the charges, e.g. the first two
non-trivial charges $\cE_4$ and $\cE_6$ are
\beaxxx
\label{I4c}
\cE_4&=&-\frac{16}{\pi^2 \cE_2}\ellE(t)\ellK(t)
-\frac{32}{\pi^4\cE_2^3}(t-1)\ellK(t)^4
\, , \\
\nonumber
\cE_6&=&\frac{2^5}{\pi^2 \cE_2}\ellE(t)\ellK(t)
+\frac{2^6}{3\pi^4\cE_2^3}\ellK(t)^3\Big[
4(t-2)\ellE(t)+7(t-1)\ellK(t)
\Big] \\
&-&\frac{2^{9}}{\pi^6\cE_2^5}(t-1)\ellK(t)^5\Big[
\ellE(t)-(t-2)\ellK(t)
\Big]
-\frac{5\cdot 2^{9}}{\pi^8\cE_2^7}(t-1)^2\ellK(t)^8\, .
\eexxxa

For the same reason as for the folded string the higher charges
$\cE_n$ evaluated on the circular type solution need to be
improved in order to scale homogeneously.
It turns out that the corresponding redefined charges are
given by the same formulae (\ref{im4}) and (\ref{im6}).
This is not surprising since
a universal relation between the string and the gauge theory models
implies that
the charge redefinition must be universal as well, {\it i.e.},
it should be independent of particular solutions of the evolution equations.

String theory provides the following values for the improved charges
at one loop:
{\small
\beaxxx
\nonumber
~~{\cal Q}_2&=&\frac{2}{\pi^2}\ellK(t_0)\ellE(t_0)\, , \\
\label{SP}
~~{\cal Q}_4&=&-\frac{16}{\pi^4}\ellK(t_0)^3\Big[(t_0-2)\ellE(t_0)
+(t_0-1)\ellK(t_0) \Big]\, , \\
\nonumber
~~{\cal Q}_6&=&\frac{2^{9}}{3\pi^6}\ellK(t_0)^5\Big[(6-6t_0+t_0^2)
\ellE(t_0)+2(2-3t_0+t_0^2)\ellK(t_0)\Big]\, , \\
\nonumber
~~{\cal Q}_8&=&
\frac{2^{11}}{\pi^8}\ellK(t_0)^7\Big[(2-t_0)(10-10t_0+t_0^2)
\ellE(t_0)-3(t_0-1)(5-5t_0+t_0^2)\ellK(t_0)\Big]\, , \\
\nonumber
{\cal Q}_{10}&=&\frac{2^{17}}{5\pi^{10}}\ellK(t_0)^9
\Big[(70-140t_0+90t_0^2-20t_0^3+t_0^4)
\ellE(t_0) \\
\nonumber
&&~~~~~~~~~~~~~~~~~~~~~~~
+4(14-35t_0+30t_0^2-10t_0^3+t_0^4)\ellK(t_0)\Big]\, \\ \nonumber
\vdots & &
\eexxxa}
and so on.
The string charge ${\cal Q}_2$ and the gauge charge
$Q_2$ in eq.(\ref{eq:Q2-2})) are related by the following
Gauss-Landen transformation
\beaxxx
\label{GLc}
t_0=-\frac{4\sqrt{r}}{(1-\sqrt{r})^2}\, .
\eexxxa
This results in
\bexxx\label{eq:transform2}
\ellK(t_0)=(1-\sqrt{r})\ellK(r),\qquad
\ellE(t_0)=2(1-\sqrt{r})^{-1} \ellE(r)-(1+\sqrt{r})\ellK(r)
\eexxx
The {\it same} transformation applies to all higher charges.
It is easy to check that under this transformation the string charges
again (\ref{SP}) nicely turn into the gauge charges
(\ref{eq:Q2-2}),(\ref{eq:Q4-2}), up to a multiplicative constant:
%
%
\beaxxx
{\cal Q}_2&=&-\frac{1}{4\pi^2}Q_2\, , ~~~
{\cal Q}_4=-\frac{3}{8\pi^4}Q_4\, , ~~~
{\cal Q}_6=-\frac{5}{12\pi^6}Q_6\, , \\
\nonumber
{\cal Q}_8&=&-\frac{7}{16\pi^8}Q_8\, ,~~~
{\cal Q}_{10}=-\frac{9}{20\pi^{10}}Q_{10}\, , \quad \ldots \quad.
\eexxxa
The general formula is
\beaxxx
\label{match2}
{\cal Q}_{2k}=-\frac{2k-1}{4k\pi^{2k}}Q_{2k}\, .
\eexxxa
Up to the factor $(-1)^{k+1}$ 
(see the comment after eq. (\ref{eq:allcharges2})) this formula coincides
with eq.(\ref{match1}).

Finally, the proof of the exact matching of the gauge and 
the ``one-loop'' string spectral curves for the circular string 
configuration proceeds in analogy to the treatment 
in chapter 4. Let us briefly sketch the main steps.
Introducing the improved spectral parameter
\beaxxx
x^2=\frac{\mu^2}{\pi^2\cJ^2} \, ,
\eexxxa
where $\mu$ is defined through eq.(\ref{imprsp}), we find from eq.(\ref{ancc})
the following relation in the large $\cJ$ limit: 
\beaxxx
x^2-\frac{1}{16\ellK(t_0)^2}\frac{z(1-z)}{1-t_0z}=0\, .
\eexxxa
The large ${\cal J}$ circular string spectral curve is then found from 
eq.(\ref{main}) to be 
\beaxxx
{\cal Q}(x)=\frac{1}{2}
\Big[
1-4x\frac{\sqrt{(1-z)(1-t_0z)}}{\sqrt{z}}\Pi(t_0z,t_0)\,
\Big].
\eexxxa
The slight difference with eq.(\ref{grbeforeGL}) 
is due to our sign convention 
for the charges (once again see the comment after eq. (\ref{eq:allcharges2})).
The Gauss-Landen transformation (\ref{GLc}) is applied by using the 
transformation rule ($x\leq c < d$)
\beaxxx
\nonumber
\Pi(t_0z,t_0)=\frac{d-c}{cd+x^2+\sqrt{(c^2-x^2)(d^2-x^2)}}
\left[ \frac{1}{8}+\frac{1}{d}\sqrt{(c^2-x^2)(d^2-x^2)}\Pi\Big(\frac{x^2}{d^2},r\Big) \right]\, ,
\eexxxa
where $c$ and $d$ are given by eqs.(\ref{eq:cd}). The result we find is
\beaxxx
{\cal Q}(x)=H(x)-\frac{\alpha}{2}\, ,
\eexxxa
where $H(x)$ is {\it precisely} the circular gauge theory spectral curve 
in eq.(\ref{eq:resolvent2}), q.e.d.

\subsection{The Gauss-Landen transformation}
It is of interest to recall the geometric meaning of the
Gauss-Landen transformation and to understand what it means in our
present context.
The Jacobi modulus $t$
completely defines an elliptic curve $y^2=(1-x^2)(1-tx^2)$, in particular,
its complex structure $\tau$
which is the ratio of the primitive periods $\omega_1$ and $\omega_2$:
$\tau=\frac{\omega_2}{\omega_1}$. The curve
can be uniformized by the function $\sn$, i.e. $y=\sn'z$, $x=\sn z$,
which has the periods $\omega_1=4\ellK(t)$ and $\omega_2=2i\ellK(1-t)$.
Thus, the complex structure is
\beaxxx
\tau=\frac{\omega_2}{\omega_1}=\frac{i}{2}\frac{\ellK(1-t)}{\ellK(t)}\, .
\eexxxa
For the string curve defining the circular
string solution in the one-loop approximation the complex
structure $\tau_s$  is therefore
\beaxxx
\tau_s=\frac{i}{2}\frac{\ellK(1-t_0)}{\ellK(t_0)}\, .
\eexxxa
Taking into account eqs.(\ref{eq:transform2}) and the transformation rule
\beaxxx
\ellK(1-t_0)=\frac{1}{2}(1-\sqrt{r})\ellK(1-r)
\eexxxa
we perform the Gauss-Landen transformation (\ref{GLc}) on $\tau_s$:
\beaxxx
\tau_s=\frac{i}{2}\frac{\ellK(1-t_0)}{\ellK(t_0)}=
\frac{i}{4}\frac{\ellK(1-r)}{\ellK(r)}=\frac{1}{2}\tau_g \, ,
\eexxxa
where $\tau_g$ appears to be the complex structure of the elliptic curve
governing the position of the Bethe roots on the gauge theory side.
We therefore realize that the gauge curve\footnote{
The curves we discussed here should not be confused with
the spectral curves, the latter depend on the additional
spectral parameter and generate the infinite family of commuting charges.}
defining the
Bethe root distribution and the string curve which solves
the string evolution equations are related in a simple way:
{\it the former is a double-cover of the latter}. A similar
interpretation might be given for the folded string solution.

The potential importance of our observation about the
interrelation between the string and the gauge curves is that the
covering index (two in our case) is an integer
and, therefore, it might be stable to loop deformations,
i.e. it could survive even if the higher loops are concerned.
This would mean that the Gauss-Landen transformation
is universal and it could be
extended to higher loops as well.

\section{Conventions for elliptic integrals and elliptic functions}

For the benefit of the serious reader we are collecting our
conventions for various elliptic integrals and functions.

The complete
elliptic integrals of the first ($\ellK$) and second ($\ellE$)
kind are
\bexxx\label{eq:conds2}
\ellK(q)\equiv \int_0^{\pi/2}\frac{d\varphi}{\sqrt{1-q\sin^2 \varphi}}
\qquad \qquad
\ellE(q)\equiv \int_0^{\pi/2} d\varphi\ \sqrt{1-q\sin^2 \varphi}.
\eexxx
and the complete elliptic integral of the third kind is
\bexxx\label{eq:third}
\Pi(m^2,q)\equiv
\int_0^{\pi/2}\frac{d\varphi}{(1-m^2 \sin^2 \varphi)
\sqrt{1-q\sin^2 \varphi}}\ .
\eexxx
To derive the string spectral curve
the following integral representation for $\Pi(m^2,q)$ was
used
\beaxxx
\int_0^{2\pi}\frac{d\sigma}{1-m^2 \sn^2 (\frac{2}{\pi} \ellK(q)\sigma,q)}=
\frac{2\pi}{\ellK(q)}\Pi(m^2,q)\, .
\eexxxa

The Jacobi elliptic function $\sn\nu\equiv \sn(\nu,q)$ inverts the elliptic integral of the
first kind:
\beaxxx
\nu = \int_0^{\sn\nu }\frac{dx}{\sqrt{(1-x^2)(1-qx^2)}} \, .
\eexxxa
The elliptic functions $\cn\nu\equiv \cn(\nu,q)$ and $\dn\nu\equiv \dn(\nu,q)$
are related to $\sn\nu$ through the identities:
\beaxxx
\sn ^2\nu+\cn^2\nu=1\, , ~~~~\dn^2\nu+q~\sn ^2\nu=1\, .
\eexxxa


\end{document}